\documentclass[conference,letter]{IEEEtran}

\usepackage[utf8]{inputenc} 
\usepackage[T1]{fontenc}
\usepackage{url}              % provides \url{...}
\usepackage{cite}             % improves presentation of citations

\usepackage[cmex10]{amsmath}  % Use the [cmex10] option to ensure compliance
\DeclareMathOperator*{\argmax}{arg\,max}

\usepackage{mleftright}       % fix to wrong spacing of \left-,
\mleftright                   % \vert dle- \right-commands 

\usepackage{graphicx}         % provides \includegraphics{...} to
\usepackage{booktabs}         % fixes poor spacing in tables and
\usepackage{soul}

\usepackage{xcolor}

\newcommand{\myVec}[1]{{\boldsymbol{#1}}}

\usepackage[font=small]{caption}

\allowdisplaybreaks[2]
\graphicspath{{figures/}}

\usepackage{amsthm}
\usepackage{amsmath}
\usepackage{amsfonts}
\usepackage{cancel}

\newtheorem{theorem}{Theorem}

\usepackage{subcaption}

\usepackage{cleveref}
\crefname{equation}{Eq}{Eqs} % capitalize "E", no period
\crefrangelabelformat{equation}{(#3#1#4--#5#2#6)}

\usepackage{amssymb}

\usepackage{float}

\usepackage{tikz}
\usetikzlibrary{backgrounds, calc, chains, fit, matrix, positioning, shadows, shapes, intersections, circuits.ee}
\tikzset{input/.style={}}
\tikzset{output/.style={}}
\tikzset{op/.style={circle, draw, fill=black!10, minimum size=2.5ex, inner sep=0ex}}
\tikzset{filter/.style={rectangle, draw, thick, fill=black!10, minimum size=3.5ex, inner sep=1ex}}
\tikzset{nn/.style={trapezium, trapezium angle=80, draw, thick, fill=black!10, inner sep=1ex}}
\tikzset{branch/.style={circle, draw, thick, fill=black, minimum size=.5ex, inner sep=0ex}}
\tikzset{tensor/.style={rectangle, draw, fill=white, minimum size=2em, double copy shadow={shadow xshift=.5ex,shadow yshift=-.5ex}}}
\tikzset{rounded/.style={rounded rectangle, draw, thick, fill=black!10, minimum size=3.5ex, inner xsep=1ex}}
\tikzset{image/.style={rectangle, draw, fill=white, minimum size=2em}}
\tikzset{>=direction ee}
\tikzset{/tikz/thin/.style={line width=.9pt}}
\tikzset{/tikz/thick/.style={line width=1.4pt}}
\tikzset{every path/.style={thin}}

\usepackage{pgfplots}
\pgfplotsset{compat=1.14}
\pgfplotsset{every axis/.append style={enlargelimits={abs=3pt},grid,axis lines=left}}
\pgfplotsset{every axis plot/.append style={thick,mark size=1.5pt,line join=bevel,mark options={solid}}}
\pgfplotsset{label style={font=\small}}
\pgfplotsset{tick label style={font=\footnotesize}}
\pgfplotsset{grid style={color=black!10}}
\pgfplotsset{legend style={draw=none,opacity=.85,font=\footnotesize,cells={anchor=west,opacity=1}}}
\pgfplotsset{every non boxed x axis/.style={xtick align=center,shorten <=-.5\pgflinewidth}}
\pgfplotsset{every non boxed y axis/.style={ytick align=center,shorten <=-.5\pgflinewidth}}
\pgfplotsset{every non boxed z axis/.style={ztick align=center,shorten <=-.5\pgflinewidth}}
\pgfplotsset{/pgf/number format/1000 sep={\,}}

\usepackage[font=small]{caption}

\IEEEoverridecommandlockouts

\begin{document}

\title{Robust Distributed Compression with \\
Learned Heegard--Berger Scheme\thanks{This work was in part done during E. Taşçı's summer internship at NYU.}} 

%%%%%%
\author{%
  \IEEEauthorblockN{Eyyüp Taşçı}
  \IEEEauthorblockA{Dept.~of Electrical and Electronics Engineering \\
  Bogazici University, Istanbul, TURKEY \\
  \texttt{eyyup.tasci@std.bogazici.edu.tr}}
\and
  \IEEEauthorblockN{Ezgi~{\"O}zyılkan, Oğuzhan Kubilay Ülger, Elza Erkip}
  \IEEEauthorblockA{Dept.~of Electrical and Computer Engineering \\
  New York University, New York, USA \\
  \texttt{\{ezgi.ozyilkan, kubi, elza\}@nyu.edu}} 
}

\maketitle 

\begin{abstract}
We consider lossy compression of an information source when decoder-only side information may be absent. This setup, also referred to as the Heegard--Berger or Kaspi problem, is a special case of robust distributed source coding. Building upon previous works on neural network-based distributed compressors developed for the decoder-only side information (Wyner--Ziv) case, we propose learning-based schemes that are amenable to the availability of side information. We find that our learned compressors mimic the achievability part of the Heegard--Berger theorem and yield interpretable results operating close to information-theoretic bounds. Depending on the availability of the side information, our neural compressors recover characteristics of the point-to-point (i.e., with no side information) and the Wyner--Ziv coding strategies that include \emph{binning} in the source space, although no structure exploiting knowledge of the source and side information was imposed into the design.
\end{abstract}

\section{Introduction} \label{sec:intro}
Imagine a distributed sensor network consisting of individual cameras positioned across various locations within a city, each independently capturing images of its surroundings. In this scenario, each sensor node compresses and sends its correlated image to a central processing unit, which then combines them to generate a comprehensive visual map of the city. However, direct communication among sensors is often infeasible, and unreliable channels (e.g., due to fading) may further hinder communication. This poses a key question: How can we leverage the correlation among sensor data in a robust manner, preventing a system failure when some nodes cannot transmit their observations? In this work, we take a first step on addressing the link failure scenario, where the decoder may not receive some correlated sensor data. 

Distributed source coding (DSC) refers to the task of efficiently compressing information from physically separated encoders. Wyner and Ziv (WZ)~\cite{Wyner:IT:76} examined a simple lossy distributed compression case where the decoder has access to a correlated source, known as the \emph{side information}, losslessly. WZ theorem relies on joint typicality and \emph{random binning} arguments, and is non-constructive. Although the theory of DSC predicts substantial improvements in compression efficiency compared to the point-to-point setups~\cite{elements_of_information_theory}, developing practical distributed compressors operating in the finite blocklength regime remains a challenging open problem to date. Recent work on learned Wyner--Ziv compressors \cite{ozyilkan2023learned,ozyilkan2023neural} showed binning-like behavior for the quadratic-Gaussian sources. Additionally, practical models employing neural networks to compress high-dimensional source data for distributed stereo image compression \cite{Mital_2022, Mital_2023} and for distributed task-aware image compression have been proposed \cite{Li_2023}.

\begin{figure}
\centering
   \includegraphics[width=0.90\linewidth]{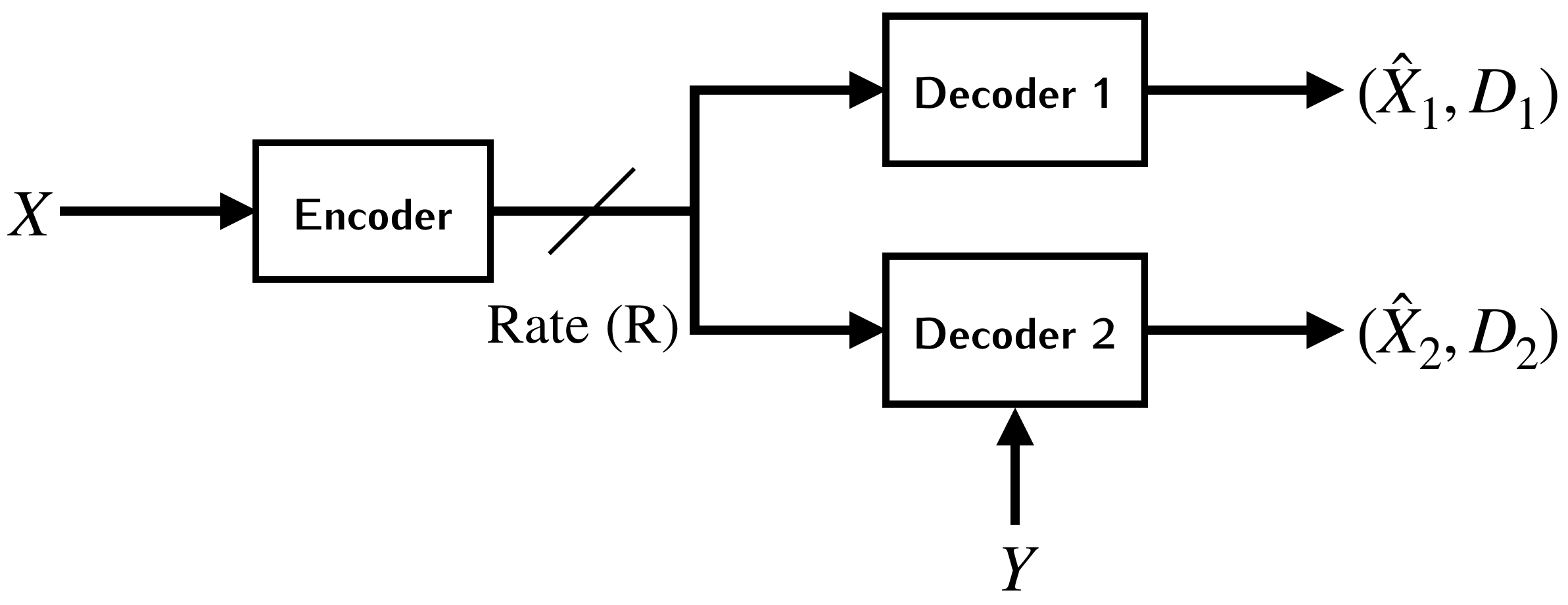}
    \caption{Lossy source coding when side information may be absent, also known as the \emph{Heegard--Berger} or the \emph{Kaspi} problem.}
    \label{fig:HB_setup} 
\end{figure}

An interesting question is whether it is still possible to achieve a reconstruction with non-trivial distortion when side information fails to reach the decoder in the WZ setup. An equivalent formulation of this problem containing two decoders, one with side information and the other without, is illustrated in Fig.~\ref{fig:HB_setup}. Heegard and Berger (HB)~\cite{Heegard_Berger85} fully characterized the asymptotic rate--distortion (R-D) function for the quadratic-Gaussian case of Fig.~\ref{fig:HB_setup}, and also considered scenarios when different side information is available at several decoders. In~\cite{Kaspi}, Kaspi also established the R-D function when side information is available at both encoder and decoder. Intuitively, one can improve the robustness of a WZ coding system against the absence of side information by reducing the number of codewords in each bin. Such an approach yields a trade-off between compression efficiency and system robustness, encapsulating the essence of the robust DSC scheme developed by Ishwar \emph{et al.}~\cite{Ishwar}. 

Recent studies in~\cite{ozyilkan2023learned, ozyilkan2023neural, ozyilkan2023neuralworkshop} have demonstrated that learning-based WZ compressors can recover different types of interpretable random binning mechanisms without any specific structure being imposed onto the design. These results offer empirical evidence that learned distributed compressors can achieve competitive constructive solutions in the non-asymptotic blocklength, closely resembling those of handcrafted frameworks such as DISCUS~\cite{DISCUS}, without requiring a priori knowledge of source statistics. In these learned WZ compressors, the encoder sends the bin index, and the decoder infers the quantization index with the help of the side information. Finally, the decoder reconstructs the source by utilizing the deduced quantization index and the side information, according to the distortion criterion. In their current form, it is not clear whether these learned WZ compressors can effectively adapt to the unavailability of side information. 

Expanding on the learned WZ compressors~\cite{ozyilkan2023learned, ozyilkan2023neural, ozyilkan2023neuralworkshop}, in this paper we find constructive solutions for the non-asymptotic regime of the HB problem where side information may be absent, by leveraging universal function approximation capabilities of artificial neural networks (ANNs)~\cite{hornik_et_al, Leshno1993}. Similar to~\cite{ozyilkan2023learned, ozyilkan2023neural, ozyilkan2023neuralworkshop}, our focus is on the one-shot regime, where each source realization is compressed one at a time, as in popular ANN-based image compressors~\cite{Balle2017, balle2018variational}. We present three unique solutions for the HB problem, where we either jointly address the quantization and binning components (Figs.~\ref{fig:unified_model} and~\ref{fig:marginal_model}) or alternatively, adopt a two-step approach that involves a learned quantizer coupled with an ideal Slepian-Wolf (SW) coder (Fig.~\ref{fig:conditional_model}). 

The paper is organized as follows. We first give an overview of HB theorem, detailing the achievability result (Section~\ref{subsec:general_setting}) and the quadratic-Gaussian setup we consider (Section~\ref{subsec:experimental_setup}). To define training objectives for the proposed robust schemes, we minimize upper bounds on mutual information. These are formulated using one of three probabilistic models employing ANNs, which can also interpreted as operational robust distributed schemes (Section~\ref{sec:neural_upper_bounds}). Finally, we discuss empirical results (Section~\ref{sec:discussion}) and conclude the paper (Section~\ref{sec:conclusion}).

\section{Heegard--Berger Problem Setup} \label{sec:hb_theorem}

\subsection{General setting} \label{subsec:general_setting} 
Let $(X,Y)$ be a pair of random variables with joint distribution $p(x,y)$ defined on the product alphabet $\mathcal{X} \times \mathcal{Y}$, where $X$ and $Y$ represent the source and decoder-only side information, respectively. As shown in Fig.~\ref{fig:HB_setup}, the encoder wishes to describe the source $X$ to two decoders, an \emph{uninformed} and an \emph{informed} one, with the latter having access to the side information $Y$. The first decoder reconstructs $\hat{X}_1$ in a point-to-point fashion while the second decoder reconstructs $\hat{X}_2$ with the help of the side information $Y$, under distortion metrics $d_i : \mathcal{X} \times \hat{\mathcal{X}}_i \rightarrow \mathbb{R}_{\geq 0}$ where $\hat{\mathcal{X}}_i$ are the reconstruction alphabets for $i\in\{1,2\}$. The goal is to find the minimum achievable rate under two expected distortion constraints: $\mathbb{E}[d_i(x,\hat{x}_i)] \leq D_i$ for some $D_i\geq 0$. In the asymptotic blocklength regime, where we consider joint compression of $n$ i.i.d. source samples as $n \rightarrow \infty$, the following theorem by Heegard and Berger~\cite{Heegard_Berger85} characterizes the optimal R-D function.

\begin{theorem} \label{theo:HB}(Heegard--Berger Theorem~\cite{Heegard_Berger85}) 
Let $(X,Y)$ be a pair of random variables with joint distribution $p(x,y)$, representing the source and the correlated side information respectively, and $d_i : \mathcal{X} \times \hat{\mathcal{X}}_i \rightarrow \mathbb{R}_{\geq 0}$ be single letter distortion measures for $i \in \{1,2\}$. The R-D function for $X$ with side information $Y$ available only at one of the decoders is:
\begin{align}
    R(D_1,D_2) = \min_{p(w,u\vert x)} (I(X;W) + I(X;U|Y, W)), \label{eq:HB}
\end{align}
where the minimization is over all conditional probability distributions $p(w, u \vert x)$ such that there exists functions $g_1(W) = \hat{X}_1$ and $g_2(W,U,Y) = \hat{X}_2$ satisfying the distortion constraints 
\begin{align}
        \mathbb{E}[d_1(X,\hat{X}_1)] \leq D_1, \; \; 
        \mathbb{E}[d_2(X,\hat{X}_2)] \leq D_2, \label{eq:distortions}
\end{align}
and $(W, U)-X-Y$ is a Markov chain.
\end{theorem}
We remark that the informed decoder encounters a WZ problem~\cite{Wyner:IT:76} (cf. second mutual information term in Eq.~\eqref{eq:HB}), while the uninformed decoder is subjected to an ordinary point-to-point R-D problem (cf. first mutual information term in Eq.~\eqref{eq:HB}). 

Note that the HB theorem considers two distortion constraints as in Eq.~\eqref{eq:distortions}. Alternatively, one can opt for a combined weighted sum distortion constraint given as:
\begin{equation}
    \mathbb{E}\left[\beta d_1(X,\hat{X}_1) + (1-\beta)d_2(X,\hat{X}_2) \right] \leq \myVec{D},
\end{equation}
for some $\beta \in [0,1]$. In this case, the asymptotically minimum weighted distortion for a fixed rate is given as~\cite{Chris_Andrea_2012}:
\begin{equation}
\label{eq:expdist}
    \begin{split}
        \myVec{D}^*(R) = &\min_{D_1, D_2: R(D_1, D_2) \leq R} \beta  D_1 + (1-\beta) D_2, \\
    \end{split}
\end{equation} 
where $R(D_1,D_2)$ is given in Theorem~\ref{theo:HB}. 
This weighted distortion measure can also be conceptualized as having a $\beta$ probability of receiving no side information (see the discussion in \cite{Chris_Andrea_2012}). As will be seen in Section~\ref{sec:neural_upper_bounds}, considering a combined distortion constraint, such as the one in Eq.~\eqref{eq:expdist}, simplifies the learning procedure of HB compressors, and yet, still offers valuable insights about how close they operate with respect to the optimum.

\subsection{Quadratic-Gaussian case} \label{subsec:experimental_setup}
Similarly to the WZ R-D function~\cite{Wyner:IT:76}, the HB formula in Eq.~\eqref{eq:HB} has a closed-form expression only in few special cases~\cite{Heegard_Berger85}. To compare the performance of our learned distributed compressors to the asymptotic HB R-D function, we consider a quadratic-Gaussian setup: Suppose $Y = X + N$, where 
$N \sim \mathrm{N}(0, \sigma_n^2)$ is independent from $X \sim \mathrm{N}(0, \sigma_x^2)$, and the distortion metrics are $d_i(x,\hat{x}_i) = (x-\hat{x}_i)^2$ for $i \in \{1,2\}$. The analytical expression of $R(D_1, D_2)$ for this case is given as \cite{Heegard_Berger85}:
\begin{equation} \label{eq:HB_Gauss}
    \begin{split}
        R(D_1,D_2) &= \frac{1}{2}\log\left(\frac{\sigma_x^2}{\Delta_1}\right) + \frac{1}{2}\log\left(\frac{\sigma_n^2\Delta_1}{\Delta_2(\Delta_1+\sigma_n^2)}\right),
    \end{split}
\end{equation}
where $\Delta_1 = \min(\sigma_x^2,D_1), \; \Delta_2 = \min([1/\Delta_1+1/\sigma_n^2]^{-1},D_2).$ The two terms of Eq.~\eqref{eq:HB_Gauss} respectively correspond to the two mutual information terms given in Theorem~\ref{theo:HB}. The first and second expressions in Eq.~\eqref{eq:HB_Gauss} represent the rate required to describe $W$ and $U$ under the distortion constraints of $D_1$ and $D_2$, respectively. There are four operationally different R-D regions depending on the value of the pair $(D_1, D_2)$. When $D_1 \leq \sigma^2_{x}$ and $D_2 \geq \frac{D_1\sigma_n^2}{D_1 + \sigma_n^2}$, the second term in Eq.~\eqref{eq:HB_Gauss} vanishes, and the problem degenerates into a standard point-to-point lossy compression problem i.e., having only the first decoder in Fig.~\ref{fig:HB_setup}. Similarly, when $D_1 \geq \sigma^2_{x}$, and $ D_2 \leq \frac{D_1\sigma_n^2}{D_1 + \sigma_n^2}$ the first term in Eq.~\eqref{eq:HB_Gauss} vanishes, and the problem degenerates to the WZ coding problem i.e., having only the second decoder in Fig.~\ref{fig:HB_setup}. When $D_1 \geq \sigma^2_{x}$, and $ D_2 \geq \frac{D_1\sigma_n^2}{D_1 + \sigma_n^2}$, both terms vanish, and the distortion constraints can be trivially satisfied with zero rate. The most operationally interesting region arises when $D_1 \leq \sigma^2_{x}$ and $ D_2 \leq \frac{D_1\sigma_n^2}{D_1 + \sigma_n^2}$. Here, both rate terms are non-zero, forcing the encoder to allocate its total rate budget between the point-to-point R-D problem of describing $W$ and the WZ R-D problem of describing $U$. In Appendix~\ref{sec:appendix_minimum_distortion}, we discuss the optimum weighted distortion outlined in Eq.~\eqref{eq:expdist} for the quadratic-Gaussian setting we consider.

\section{Operational Neural Heegard--Berger Schemes}
\label{sec:neural_upper_bounds}

\begin{figure}
\centering
\begin{subfigure}[b]{\columnwidth}
   \includegraphics[width=1.02\linewidth]{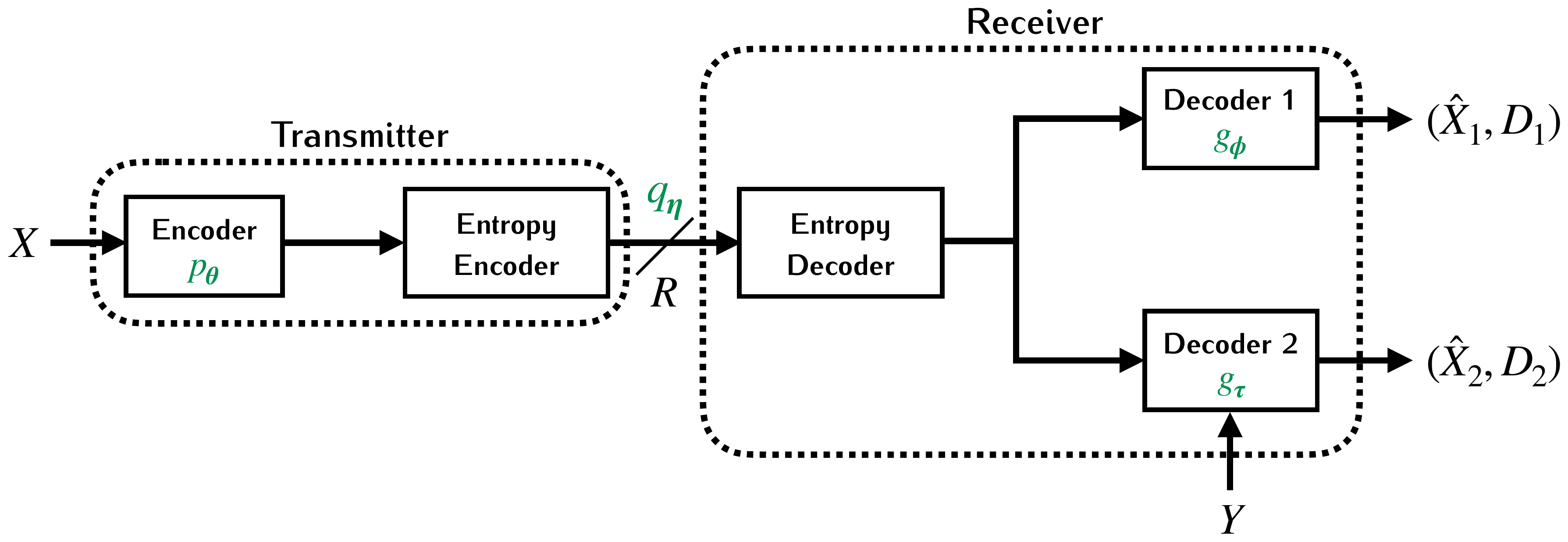}
   \caption{}
    \label{fig:unified_model} 
\end{subfigure}
\begin{subfigure}[b]{\columnwidth}
   \includegraphics[width=1.02\linewidth]{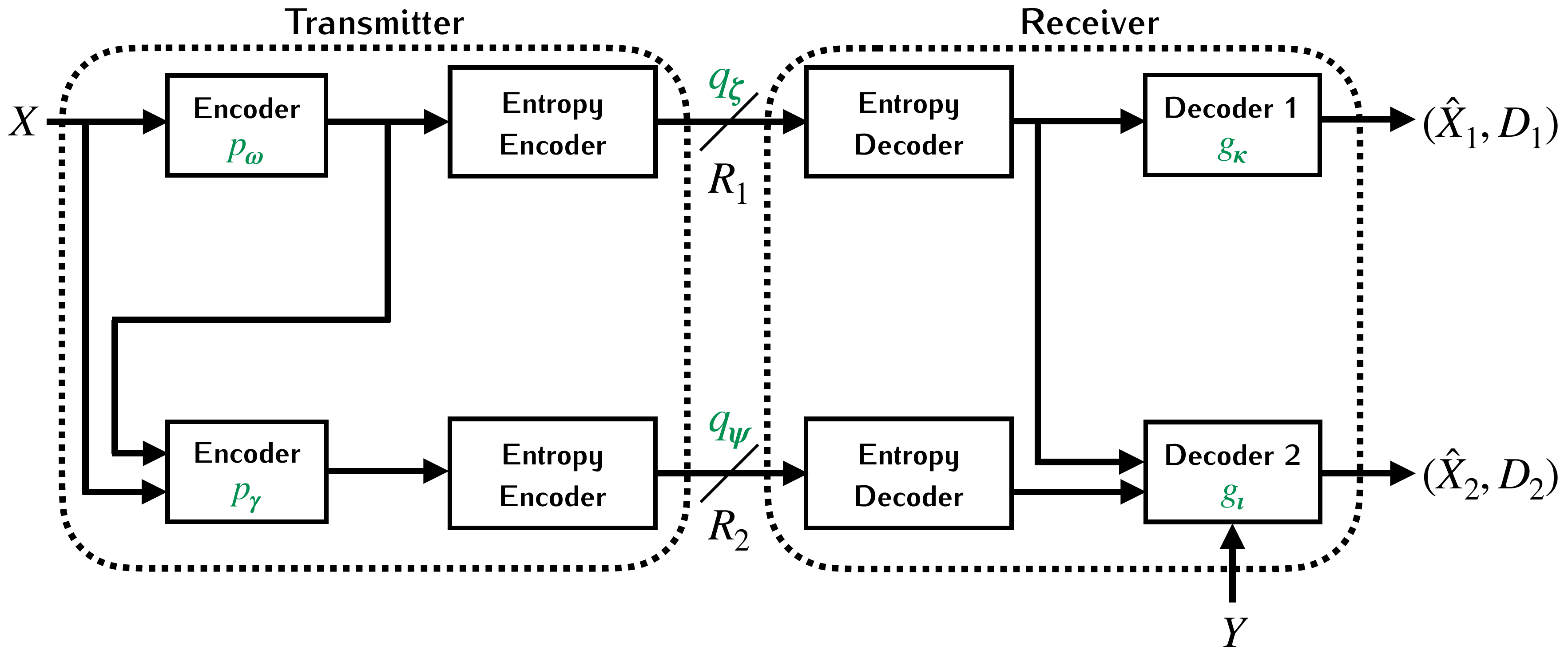}
   \caption{}
   \label{fig:marginal_model} 
\end{subfigure}
\begin{subfigure}[b]{\columnwidth}
   \includegraphics[width=1.02\linewidth]{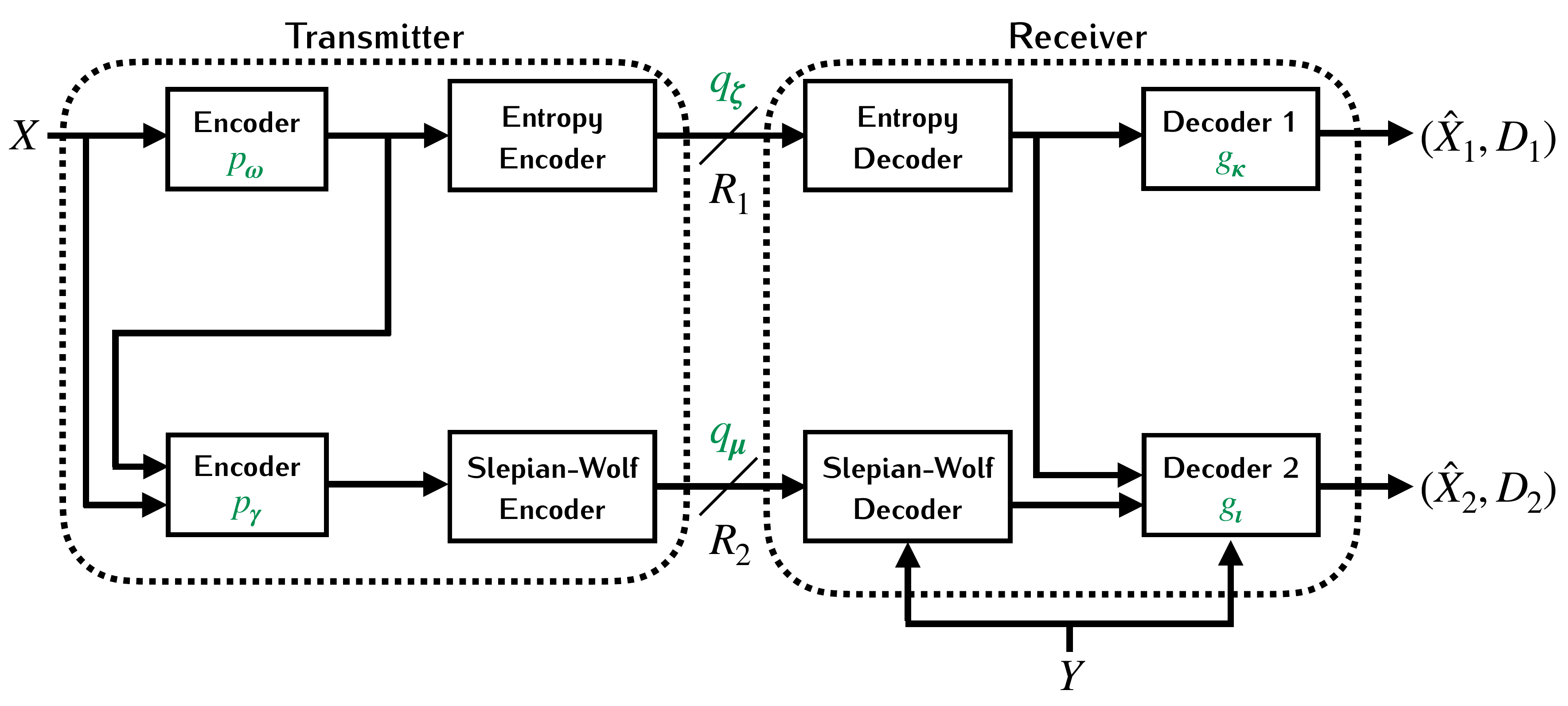}
   \caption{}
   \label{fig:conditional_model} 
\end{subfigure}
\caption{The three lossy compression systems that we consider: (a) learned compressor sending a joint description for both decoders using a classic entropy coder (i.e., the \emph{joint} formulation, see Eq.~\eqref{eq:proposed_loss_joint}), (b) learned compressors sending individual descriptions to both decoders using classic entropy coders (i.e., the \emph{marginal} formulation, see Eq.~\eqref{eq:proposed_loss_marginal}), and (c) using a combination of a classic entropy coder and an ideal Slepian--Wolf coder (i.e., the \emph{conditional} formulation, see Eq.~\eqref{eq:proposed_loss_conditional}). The learned parameters are indicated in green.}
\label{fig:sys}
\end{figure}

For our learned compressors, we first consider the system model in Fig.~\ref{fig:unified_model}, which corresponds to a straightforward parametrization of the HB setup depicted in Fig.~\ref{fig:HB_setup}. Note that in this case, a joint description is sent to both uninformed and informed decoders, the latter having access to the side information $Y$. By introducing the auxiliary variable $V=(W, U)$, we first assume that the encoder in the achievability proof of HB theorem can be represented by a probability model $p_{\boldsymbol{\theta}}(v|x)$ with parameters $\boldsymbol{\theta}$. This formulation, which we name as \emph{joint}, yields the following variational upper bound:
\begin{align}
 I(X;W) + I(X;U|Y, W)  \leq 
   & I(X;U,W) = I(X;V), \label{eq:joint_upper_bound} \\
   &\leq \mathbb{E}_{\substack{p(x)\\p_{\boldsymbol{\theta}}(v \vert x)}} \bigg[ \log \frac{p_{\boldsymbol{\theta}}(v \vert  x)}{{q_{\boldsymbol{\eta}}(v)}} \bigg], \label{eq:upper_bound_merg_2} \\ 
    &= \mathbb{E}_{\substack{p(x)}} \; \mathrm{r}_\mathrm{j}(x), \label{eq:upper_bound_merg}
\end{align}
where $q_{\boldsymbol{\eta}}(v)$, with parameters $\boldsymbol{\eta}$, is a model of the distribution $p(v) \triangleq p(w, u)$, which is generally not known in closed form. $\mathrm{r}_\mathrm{j}(x)$ refers to the expectation in Eq.~(\ref{eq:upper_bound_merg_2}) with respect to $p_{\boldsymbol{\theta}}(v \vert x)$. The proof of the upper bound in Eq.~\eqref{eq:joint_upper_bound} can be found in Appendix~\ref{sec:appendix_upper_bound}.
The upper bound in Eq.~\eqref{eq:upper_bound_merg} follows from cross-entropy~\cite{kullback1997information} being larger or equal to entropy~\cite{elements_of_information_theory}.

Next, we consider the system models illustrated in Figs.~\ref{fig:marginal_model} and~\ref{fig:conditional_model}. In these schemes, we opt for a layered encoding approach, aligned more closely with the HB theorem, where we separately encode the auxiliary variables $W$ and $U$ with probability models $p_{\boldsymbol{\omega}}(w|x)$ and $p_{\boldsymbol{\gamma}}(u|w, x)$ (with parameters $\boldsymbol{\omega}$ and $\boldsymbol{\gamma}$), respectively. For our objective functions, building onto Theorem~\ref{theo:HB}, we will consider either a \emph{marginal} or a \emph{conditional} formulation~\cite{ozyilkan2023learned} whose variational upper bounds respectively are:
\begin{align}
\label{eq:upper_bound_sep_marg}
    I(X;W) &+ I(X;U|Y, W) 
     \leq \mathbb{E}_{\substack{p(x)\\p_{\boldsymbol{\omega}}(w\vert x)\\p_{\boldsymbol{\gamma}}(u\vert w, x)}} \bigg[\log \frac{p_{\boldsymbol{\omega}}(w \vert  x)}{{q_{\boldsymbol{\zeta}}(w)}} \nonumber \\
     & \hspace{.5cm} + \log \frac{p_{\boldsymbol{\gamma}}(u \vert  w, x)}{q_{\boldsymbol{\psi}}{(u \vert w)}} \bigg] = \mathbb{E}_{\substack{p(x)}} \;  \mathrm{r}_\mathrm{m}(x),
\end{align}
\begin{align}
\label{eq:upper_bound_sep_cond}
    I(X;W) &+ I(X;U|Y, W) 
     \leq \mathbb{E}_{\substack{p(x,y)\\p_{\boldsymbol{\omega}}(w\vert x)\\p_{\boldsymbol{\gamma}}(u\vert w, x)}} \bigg[ \log \frac{p_{\boldsymbol{\omega}}(w \vert  x)}{{q_{\boldsymbol{\zeta}}(w)}} \nonumber \\
     & \hspace{.5cm} + \log \frac{p_{\boldsymbol{\gamma}}(u \vert  w, x)}{q_{\boldsymbol{\mu}}{(u \vert w, y)}} \bigg] = \mathbb{E}_{\substack{p(x,y)}} \; \mathrm{r}_\mathrm{c}(x, y).
\end{align}
where $\mathrm{r}_{\mathrm{m}}(x)$ and $\mathrm{r}_{\mathrm{c}}(x,y)$ refer to respective expectations in Eqs.~\eqref{eq:upper_bound_sep_marg} and \eqref{eq:upper_bound_sep_cond} with respect to $p_{\boldsymbol{\omega}}(w\vert x)$ and $p_{\boldsymbol{\gamma}}(u \vert  w, x)$. Here $q_{\boldsymbol{\zeta}}(w)$ (with parameters $\boldsymbol{\zeta}$) is a model of the distribution $p(w)$, and $q_{\boldsymbol{\psi}}(u \vert w)$ and $q_{\boldsymbol{\mu}}(u \vert w, y)$ (with parameters $\boldsymbol{\psi}$ and $\boldsymbol{\mu}$, respectively) are two different models of the distribution $p(u \vert w, y)$.

\begin{figure*}[t]
\centering
\begin{subfigure}{.5\textwidth}
  \centering
  \includegraphics[width=0.80\linewidth]{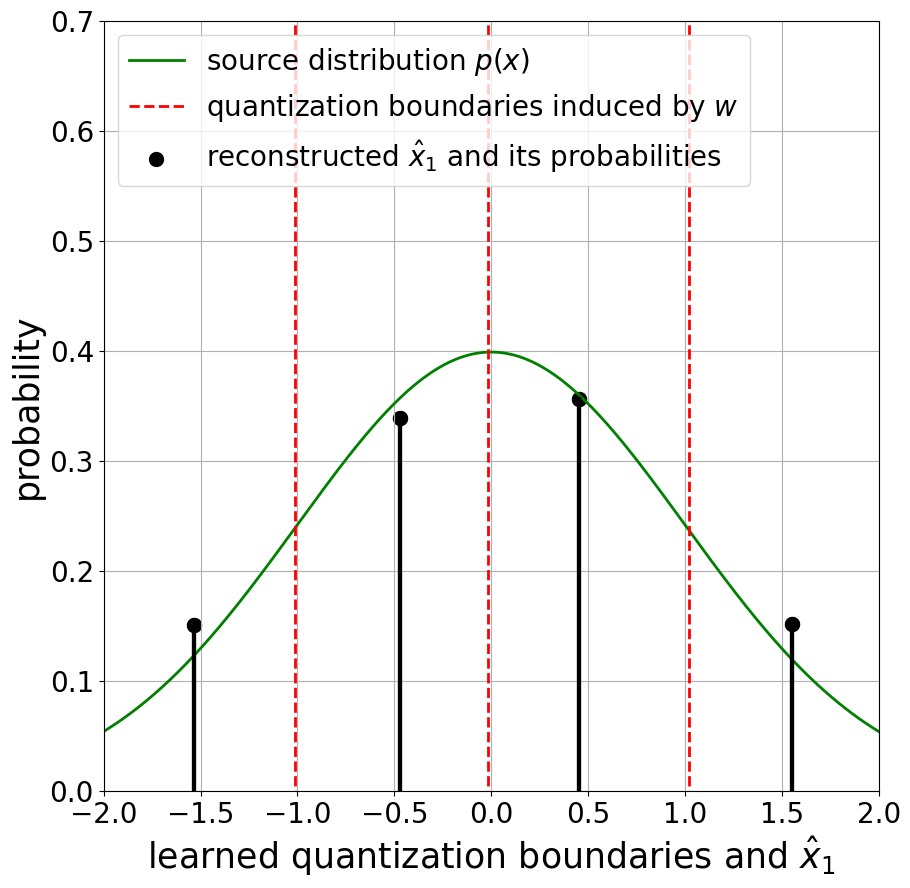}
\end{subfigure}%
\hfill
\begin{subfigure}{.5\textwidth}
  \centering
  \includegraphics[width=0.9\linewidth]{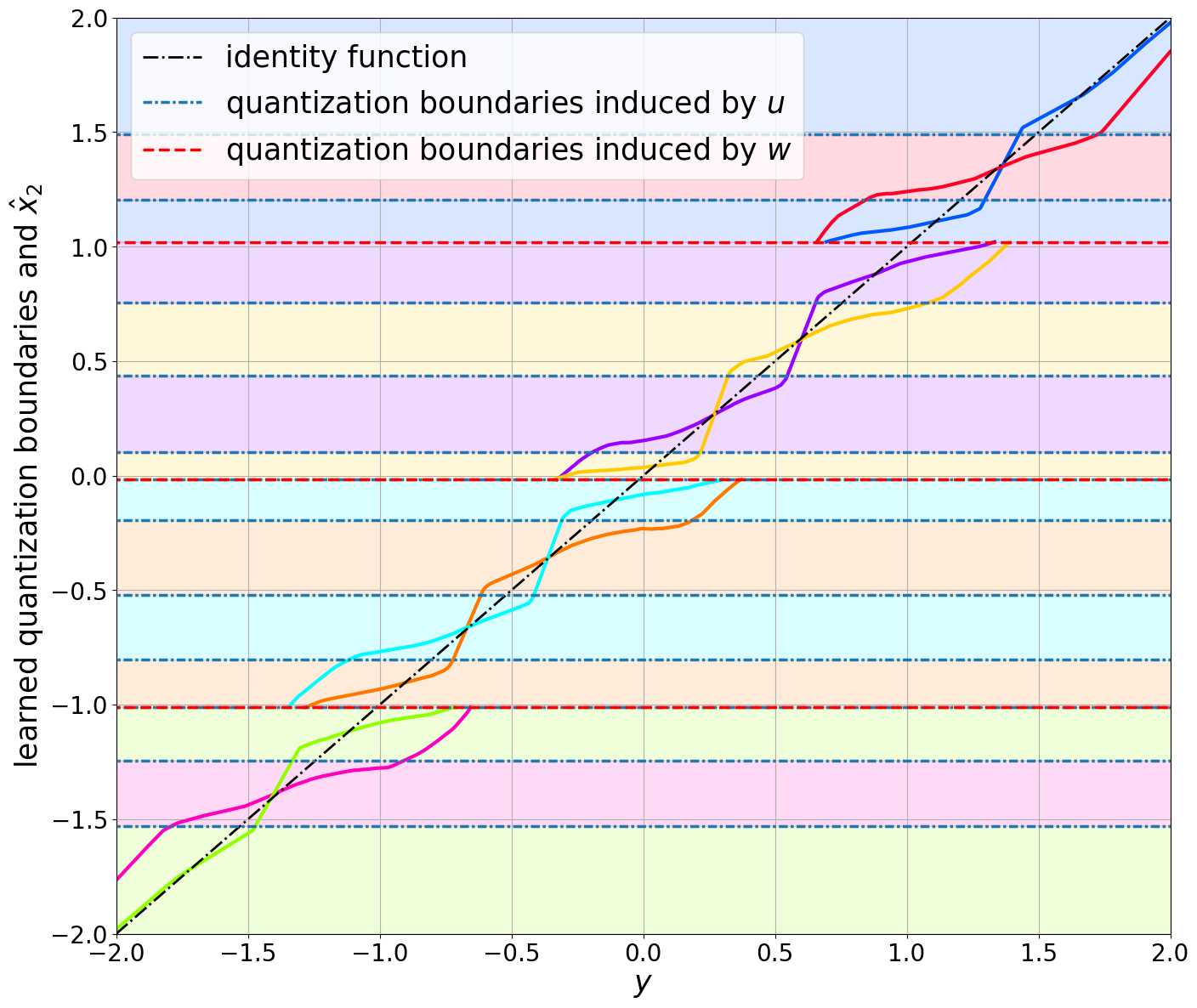}
\end{subfigure}
\caption{Visualizations of the learned optimized encoders,  $w = \argmax_{h} p_\myVec{\omega}(h|x)$ and $u = \argmax_{l} p_\myVec{\gamma}(l|w, x)$ in Eq.~\eqref{eq:upper_bound_sep_marg}, and of the decoders, $\hat{x}_{1} = g_\myVec{\kappa}(w)$ and $\hat{x}_{2}= g_\myVec{\iota}(w, u, y)$ in Eq~\eqref{eq:d_s}, for the marginal formulation (see Fig.~\ref{fig:marginal_model} and Eq.~\eqref{eq:proposed_loss_marginal}). \textbf{Left:} The dashed vertical red lines are quantization boundaries induced by $w$, and the codebook points represent the outputs of the decoder $g_\myVec{\kappa}$ to which all source values within the corresponding quantization region are mapped. The height of each codebook stalk represents the likelihood of that code vector under the entropy model $q_{\myVec{\zeta}}$ (see Eq.~\eqref{eq:upper_bound_sep_marg}). \textbf{Right:} The dashed horizontal red lines coincide with those depicted in the left panel induced by $w$, while the dotted-dashed horizontal blue lines are quantization boundaries induced by $u$. The colors between each boundary represent a unique $(w,u)$ pair. We depict the decoding function learned by $g_\myVec{\iota}$ with the solid lines, each representing a different pair of $(w,u)$ as inputs within its respective quantization region, distinguished by unique colors. The experimental setup parameters (see Section~\ref{subsec:experimental_setup}) are configured as $\sigma^{2}_x=1.00$, $\sigma^2_n = 0.01$ and $\beta = 0.20$. This model attains -15.55 dB at a rate of 2.85 bits (see Fig.~\ref{fig:RD_var_0.1} in Appendix~\ref{sec:appendix_experiments}).
}
\vspace{-0.5cm}
\label{fig:visualizations}
\end{figure*}

We will utilize these probabilistic models (i.e., $p_{\boldsymbol{\theta}}(v \vert x), p_{\boldsymbol{\omega}}(w\vert x)$ and $p_{\boldsymbol{\gamma}}(u \vert w, x)$) to aid in the learning process of respective encoders. We set the encoder outputs in a deterministic way, similar to~\cite{ozyilkan2023learned}, e.g., $v = \argmax_{h}{p_{\boldsymbol{\theta}}(h|x)}$. To actualize a practice-oriented compression setting, we also set all encoder outputs (i.e., $V$, $W$, and $U$) as discrete. 
Similarly to~\cite{ozyilkan2023learned, ozyilkan2023neuralworkshop, ozyilkan2023neural}, without loss of generality, we define all probabilistic models as discrete distributions with probabilities $P_k = \frac{\exp \alpha_k}{\sum_{i=1}^K\exp \alpha_i }$ for $k \in \{1, \dots, K\}$, where $K$ is a model parameter. The unnormalized log-probabilities (\emph{logits}) are either directly treated as learnable parameters (i.e., $q_{\boldsymbol{\eta}}(v)$ and $q_{\boldsymbol{\zeta}}(w)$), or computed by ANNs as functions of the conditioning variable (e.g., $w$ and $y$ for $q_{\boldsymbol{\mu}}{(u \vert w, y)}$). These design choices keep the parametric families as general as possible without imposing any structure. Specifically, this allows the encoders to learn, when needed, quantization schemes that may incorporate discontiguous bins in the source space, resembling the random binning operation in the achievability part of the HB and WZ theorems. Previously, it was demonstrated in~\cite{ozyilkan2023learned, ozyilkan2023neuralworkshop, ozyilkan2023neural} that a similarly parametrized neural distributed compressor can learn different types of interpretable binning mechanisms for the WZ problem, such as periodic-like mappings for the quadratic-Gaussian case.

\vspace{-0.1cm}
Since we focus on a weighted distortion metric (see Eq.~\eqref{eq:expdist}), instead of having individual distortion constraints as is the case for the HB theorem (see Section~\ref{sec:hb_theorem}), we define the following distortion functions for the cases where there is either joint (see Fig.~\ref{fig:unified_model}) or separate descriptions (see Fig.~\ref{fig:marginal_model} and Fig.~\ref{fig:conditional_model}) to the decoders, respectively, as:
 \begin{align} 
    \mathrm{d}_\mathrm{j}(x,y) &= \beta d(x, g_{\boldsymbol{\phi}}(v)) + (1-\beta) d(x, g_{\boldsymbol{\tau}}(v, y)),  \label{eq:d_u}  \\
    \mathrm{d}_\mathrm{s}(x,y) &= \beta d(x, g_{\boldsymbol{\kappa}}(w)) + (1-\beta)d(x, g_{\boldsymbol{\iota}}(u, w, y)), \label{eq:d_s}
\end{align} 
where $g_{\boldsymbol{\phi}}(v)$ and $g_{\boldsymbol{\kappa}}(w)$ denote the uninformed decoders, represented by ANNs with parameters $\boldsymbol{\phi}$ and $\boldsymbol{\kappa}$ respectively, which yield $\hat{x}_{1}$. Similarly, $g_{\boldsymbol{\tau}}(v, y)$ and $g_{\boldsymbol{\iota}}(u, w, y)$ correspond to the informed decoders, represented by ANNs with parameters $\boldsymbol{\tau}$ and $\boldsymbol{\iota}$ respectively, which output $\hat{x}_{2}$. Building on the upper bounds developed in Eqs.~\eqref{eq:upper_bound_merg},~\eqref{eq:upper_bound_sep_marg} and~\eqref{eq:upper_bound_sep_cond} in tandem with the distortion metrics defined in Eqs.~\eqref{eq:d_u} and~\eqref{eq:d_s}, this yields the loss functions for three different variants we consider:
\begin{align} \label{eq:proposed_loss_joint}
     &L_\mathrm{j}(\boldsymbol{\theta}, \boldsymbol{\eta}, \boldsymbol{\phi}, \boldsymbol{\tau}) = \mathbb{E} [\mathrm{r}_\mathrm{j}(x) + \lambda \cdot  \mathrm{d}_\mathrm{j}(x, y) ], \\ \label{eq:proposed_loss_marginal}
     &L_\mathrm{m}(\boldsymbol{\omega}, \boldsymbol{\gamma}, \boldsymbol{\zeta}, \boldsymbol{\psi}, \boldsymbol{\kappa}, \boldsymbol{\iota}) = \mathbb{E} [\mathrm{r}_\mathrm{m}(x) + \lambda \cdot  \mathrm{d}_\mathrm{s}(x, y) ], \\ \label{eq:proposed_loss_conditional}
     &L_\mathrm{c}(\boldsymbol{\omega}, \boldsymbol{\gamma}, \boldsymbol{\zeta}, \boldsymbol{\mu}, \boldsymbol{\kappa}, \boldsymbol{\iota}) = \mathbb{E} [\mathrm{r}_\mathrm{c}(x, y) + \lambda \cdot  \mathrm{d}_\mathrm{s}(x, y) ],
\end{align} 
where we relax the constrained formulation of the HB theorem to unconstrained ones using Lagrange multipliers. Here, $\{\boldsymbol{\theta}, \boldsymbol{\eta}, \boldsymbol{\phi}, \boldsymbol{\tau}, \boldsymbol{\omega}, \boldsymbol{\gamma}, \boldsymbol{\zeta}, \boldsymbol{\psi}, \boldsymbol{\mu}, \boldsymbol{\kappa}, \boldsymbol{\iota}\}$ are optimization parameters, and $\lambda > 0$ controls the trade-off. We can obtain different points in the achievable R-D region induced by the weighted distortion constraint by simply varying the parameter $\lambda$. The optimized $p_{\boldsymbol{\theta}}(v \vert x), p_{\boldsymbol{\omega}}(w \vert x)$ and $ p_{\boldsymbol{\gamma}}(u \vert w, x)$ models  yield the ANN-based encoders depicted in Fig.~\ref{fig:sys}. Similarly, the optimized $\big(g_{\boldsymbol{\phi}}(v), g_{\boldsymbol{\tau}}(v, y)\big)$ and $\big(g_{\boldsymbol{\kappa}}(w), g_{\boldsymbol{\iota}}(u, w, y)\big)$ correspond to a pair of uninformed-informed decoder components for joint and separate description cases, respectively. We discuss the operational meaning of these different schemes in Appendix~\ref{sec:appendix_operational}.
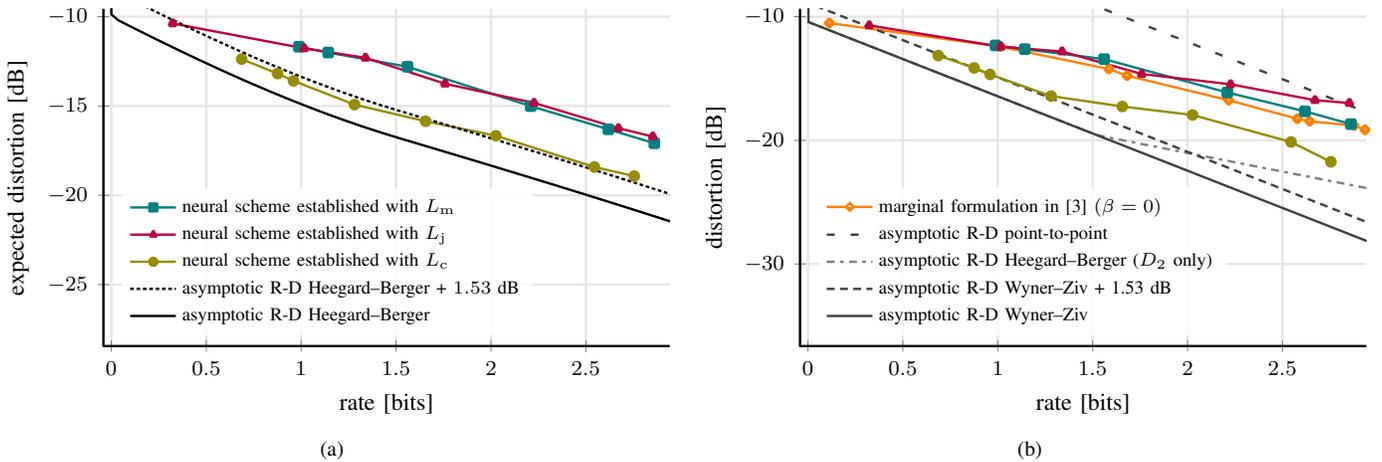
\begin{figure*}[ht]
\begin{subfigure}{\columnwidth}
    \raggedleft
    \begin{tikzpicture}[trim axis right]
    \begin{axis}[
      height=.19\textheight,
      width=.85\linewidth,
      scale only axis,
      xlabel={rate [bits]},
      ylabel={expected distortion [dB]},
      xmin=0.,
      xmax=2.9,
      ymin=-28.,
      ymax=-10.,
      legend pos=south west,
      legend style={font=\scriptsize},
      ]
      \addplot[color=teal, mark=square*] table {
0.9884 -11.681302257194982
1.1426 -12.006594505464182
1.5601 -12.798406965940432
2.209558 -15.005878743277245
2.618692 -16.303945659156497
2.860794 -17.08608946921593
      };
       \addplot[color=purple,mark=triangle*] table {
0.3228 -10.376306643299788
1.016357 -11.771457022039735
1.338725 -12.298773318547449
1.758811 -13.75387924487024
2.227076 -14.825797525246283
2.671622 -16.24372959745979
2.852585 -16.708869973181805
      };
      \addplot[color=olive,mark=*] table {
0.6866 -12.373214362725637
0.875706 -13.1826141841
0.959458 -13.5945893859
1.2812 -14.921441283041691
1.655841 -15.8534413709
2.026015 -16.6663334449
2.545 -18.416375079047505
2.755 -18.927900303521316
      };
      \addplot[color=black,densely dotted] table {
4.97571560626259 -28.47
4.919571331038443 -28.14
4.861367881963664 -27.81
4.8010960947652865 -27.47
4.738771705428015 -27.15
4.67441558488257 -26.82
4.608054471839927 -26.49
4.539721042959324 -26.16
4.469450208268615 -25.83
4.397284279195515 -25.50
4.323268651704019 -25.17
4.247451084218616 -24.84
4.169883519218755 -24.51
4.090619851219479 -24.18
4.0097156224636885 -23.85
3.9272313538130823 -23.52
3.843227749899218 -23.19
3.7577603129253374 -22.86
3.6708911381301688 -22.53
3.582684004297858 -22.20
3.4931974734089386 -21.87
3.402494232818765 -21.54
3.3106267691877393 -21.21
3.217660652239984 -20.88
3.123648273462121 -20.55
3.0286449599932506 -20.22
2.932708428441499 -19.89
2.8358849793464285 -19.56
2.7382299069287828 -19.23
2.6397831429336494 -18.9
2.5405996013158436 -18.57
2.4407206728576765 -18.24
2.3401872660177863 -17.91
2.239039598938222 -17.58
2.137317020732478 -17.25
2.0350578613209733 -16.92
1.9322936710645038 -16.59
1.8290490673616713 -16.26
1.7253712083563386 -15.93
1.6212758063840746 -15.6
1.516814011509681 -15.27
1.4160193467144244 -14.94
1.3210576009557544 -14.61
1.2309291140570155 -14.28
1.1448602144524138 -13.95
1.062240058227663 -13.62
0.982578110241563 -13.29
0.9054746653834899 -12.96
0.8305998861656883 -12.63
0.7576785641180476 -12.3
0.6864788256089763 -11.97
0.6168036164033258 -11.64
0.5484841823856696 -11.31
0.48137500952467316 -10.98
0.41534984745017406 -10.65
0.3502985492076899 -10.32
0.2861245337382722 -9.99
0.2227427291125361 -9.66
0.16007789094459102 -9.33
0.09806321652120281 -9.0
0.03663919416342865 -8.67
0.0 -8.340000000000002
0.0 4.529999999999999
      };
      \addplot[color=black] table {
4.97571560626259 -30.0
4.919571331038443 -29.67
4.861367881963664 -29.340000000000003
4.8010960947652865 -29.009999999999998
4.738771705428015 -28.68
4.67441558488257 -28.35
4.608054471839927 -28.02
4.539721042959324 -27.69
4.469450208268615 -27.36
4.397284279195515 -27.029999999999998
4.323268651704019 -26.7
4.247451084218616 -26.37
4.169883519218755 -26.04
4.090619851219479 -25.709999999999997
4.0097156224636885 -25.38
3.9272313538130823 -25.049999999999997
3.843227749899218 -24.72
3.7577603129253374 -24.39
3.6708911381301688 -24.059999999999995
3.582684004297858 -23.730000000000004
3.4931974734089386 -23.4
3.402494232818765 -23.07
3.3106267691877393 -22.740000000000002
3.217660652239984 -22.41
3.123648273462121 -22.080000000000002
3.0286449599932506 -21.75
2.932708428441499 -21.419999999999998
2.8358849793464285 -21.09
2.7382299069287828 -20.76
2.6397831429336494 -20.43
2.5405996013158436 -20.099999999999998
2.4407206728576765 -19.77
2.3401872660177863 -19.439999999999998
2.239039598938222 -19.11
2.137317020732478 -18.779999999999998
2.0350578613209733 -18.45
1.9322936710645038 -18.119999999999997
1.8290490673616713 -17.79
1.7253712083563386 -17.46
1.6212758063840746 -17.13
1.516814011509681 -16.8
1.4160193467144244 -16.47
1.3210576009557544 -16.14
1.2309291140570155 -15.809999999999999
1.1448602144524138 -15.48
1.062240058227663 -15.149999999999999
0.982578110241563 -14.82
0.9054746653834899 -14.489999999999998
0.8305998861656883 -14.16
0.7576785641180476 -13.83
0.6864788256089763 -13.499999999999998
0.6168036164033258 -13.17
0.5484841823856696 -12.839999999999998
0.48137500952467316 -12.509999999999998
0.41534984745017406 -12.18
0.3502985492076899 -11.849999999999998
0.2861245337382722 -11.52
0.2227427291125361 -11.19
0.16007789094459102 -10.86
0.09806321652120281 -10.53
0.03663919416342865 -10.2
0.0 -9.870000000000001
0.0 2.999999999999999
      };
      \legend{neural scheme established with $L_{\mathrm{m}}$, neural scheme established with $L_{\mathrm{j}}$, neural scheme established with $L_{\mathrm{c}}$, asymptotic R-D Heegard--Berger + $1.53$ dB, asymptotic R-D Heegard--Berger}
    \end{axis}
    \end{tikzpicture}
  \caption{}
  \label{subfig:RD_var0.1}
\end{subfigure}%
\hfill%
\begin{subfigure}{\columnwidth}
    \raggedleft
    \begin{tikzpicture}[trim axis right]
    \begin{axis}[
      height=.19\textheight,
      width=.85\linewidth,
      scale only axis,
      xlabel={rate [bits]},
      ylabel={distortion [dB]},
      xmin=0.,
      xmax=2.9,
      ymin=-36,
      ymax=-10.,
      legend pos=south west,
      legend style={font=\scriptsize},
      ]
            \addplot[color=orange,mark=halfsquare*] table {
0.1132 -10.510982390297862
1.014758 -12.423079622336374
1.585004 -14.230370285143108
1.681597 -14.76747921975306
2.217435 -16.757381278669314
2.578578 -18.253588073395516
2.643272 -18.456977803153357
2.87772 -18.807111108753926
2.935269 -19.146737375760114
      };
      \addplot[color=darkgray,loosely dashed] table {
      4.9828921423310435 -30.0
      0.049828921423310135 -0.2999999999999981
      0.0 0.030000000000001026
      0.0 2.999999999999999
      };
      \addplot[color=gray, dashdotted] table {
0	-10.4139268515823
0.0505050505050505	-10.7179975542731
0.101010101010101	-11.0220682569640
0.151515151515152	-11.3261389596549
0.202020202020202	-11.6302096623458
0.252525252525253	-11.9342803650367
0.303030303030303	-12.2383510677276
0.353535353535354	-12.5424217704185
0.404040404040404	-12.8464924731094
0.454545454545455	-13.1505631758003
0.505050505050505	-13.4546338784912
0.555555555555556	-13.7587045811820
0.606060606060606	-14.0627752838729
0.656565656565657	-14.3668459865638
0.707070707070707	-14.6709166892547
0.757575757575758	-14.9749873919456
0.808080808080808	-15.2790580946365
0.858585858585859	-15.5831287973274
0.909090909090909	-15.8871995000183
0.959595959595960	-16.1912702027092
1.01010101010101	-16.4953409054001
1.06060606060606	-16.7994116080909
1.11111111111111	-17.1034823107818
1.16161616161616	-17.4075530134727
1.21212121212121	-17.7116237161636
1.26262626262626	-18.0156944188545
1.31313131313131	-18.3197651215454
1.36363636363636	-18.6238358242363
1.41414141414141	-18.9279065269272
1.46464646464646	-19.2319772296181
1.51515151515152	-19.5360479323090
1.56565656565657	-19.7130958917088
1.61616161616162	-19.8651312430542
1.66666666666667	-20.0171665943997
1.71717171717172	-20.1692019457451
1.76767676767677	-20.3212372970906
1.81818181818182	-20.4732726484360
1.86868686868687	-20.6253079997815
1.91919191919192	-20.7773433511269
1.96969696969697	-20.9293787024724
2.02020202020202	-21.0814140538178
2.07070707070707	-21.2334494051633
2.12121212121212	-21.3854847565087
2.17171717171717	-21.5375201078541
2.22222222222222	-21.6895554591996
2.27272727272727	-21.8415908105450
2.32323232323232	-21.9936261618905
2.37373737373737	-22.1456615132359
2.42424242424242	-22.2976968645814
2.47474747474748	-22.4497322159268
2.52525252525253	-22.6017675672723
2.57575757575758	-22.7538029186177
2.62626262626263	-22.9058382699631
2.67676767676768	-23.0578736213086
2.72727272727273	-23.2099089726540
2.77777777777778	-23.3619443239995
2.82828282828283	-23.5139796753449
2.87878787878788	-23.6660150266904
2.92929292929293	-23.8180503780358
2.97979797979798	-23.9700857293813
3.03030303030303	-24.1221210807267
3.08080808080808	-24.2741564320722
3.13131313131313	-24.4261917834176
3.18181818181818	-24.5782271347630
3.23232323232323	-24.7302624861085
3.28282828282828	-24.8822978374539
3.33333333333333	-25.0343331887994
3.38383838383838	-25.1863685401448
3.43434343434343	-25.3384038914903
3.48484848484849	-25.4904392428357
3.53535353535354	-25.6424745941812
3.58585858585859	-25.7945099455266
3.63636363636364	-25.9465452968720
3.68686868686869	-26.0985806482175
3.73737373737374	-26.2506159995629
3.78787878787879	-26.4026513509084
3.83838383838384	-26.5546867022538
3.88888888888889	-26.7067220535993
3.93939393939394	-26.8587574049447
3.98989898989899	-27.0107927562902
4.04040404040404	-27.1628281076356
4.09090909090909	-27.3148634589810
4.14141414141414	-27.4668988103265
4.19191919191919	-27.6189341616719
4.24242424242424	-27.7709695130174
4.29292929292929	-27.9230048643628
4.34343434343434	-28.0750402157083
4.39393939393939	-28.2270755670537
4.44444444444445	-28.3791109183992
4.49494949494950	-28.5311462697446
4.54545454545455	-28.6831816210901
4.59595959595960	-28.8352169724355
4.64646464646465	-28.9872523237809
4.69696969696970	-29.1392876751264
4.74747474747475	-29.2913230264718
4.79797979797980	-29.4433583778173
4.84848484848485	-29.5953937291627
4.89898989898990	-29.7474290805082
4.94949494949495	-29.8994644318536
5	-30.1452053875792
      };
      \addplot[color=darkgray, densely dashed] table {
      3.253176333012395 -28.47
      0.019279332639547557 -9.0
      0.0 -8.67
      };
      \addplot[color=darkgray] table {
        3.253176333012395 -30.0
        0.002835788569855034 -10.431
        0.0 -4.359000000000002
      };
    \addplot[color=teal, mark=square*] table {
0.9884 -12.321023839819095
1.1426 -12.620126736665693
1.5601 -13.42944147142896
2.209558 -16.127347842158247
2.618692 -17.654574086884715
2.860794 -18.68798509973767
         };
     \addplot[color=purple,mark=triangle*] table {
0.3228 -10.700704399154121
1.016357 -12.426267494869307
1.338725 -12.813322646837893
1.758811 -14.633027063918151
2.227076 -15.466052063867226
2.671622 -16.749361064351138
2.852585 -16.979072841566442
      };
      \addplot[color=olive,mark=*] table {
0.6866 -13.142582613977362
0.875706 -14.1402068293
0.959458 -14.6671104199
1.2812 -16.420651529995464
1.655841 -17.26165682
2.026015 -17.947144165
2.545 -20.13228265733755
2.755 -21.739251972991735
      };
      \legend{marginal formulation in~\cite{ozyilkan2023learned} $(\beta = 0)$, asymptotic R-D point-to-point, asymptotic R-D Heegard--Berger ($D_2$ only), asymptotic R-D Wyner--Ziv + 1.53 dB, asymptotic R-D Wyner--Ziv};
    \end{axis}
    \end{tikzpicture}
    \caption{}
    \label{subfig:RD_ablation_var_0.1}
\end{subfigure}%
\caption{Rate--distortion (R-D) performances obtained with joint, marginal and conditional formulations (see Eqs.~\eqref{eq:proposed_loss_joint},~\eqref{eq:proposed_loss_marginal} and~\eqref{eq:proposed_loss_conditional}), where experimental setup parameters (see Section~\ref{subsec:experimental_setup}) are set as $\sigma^{2}_{x}=1.00$, $\sigma^{2}_{n} = 0.10$ and $\beta = 0.01$. In both panels, we plot the empirical results versus the asymptotic bounds. We provide the expected distortion achieved by two decoders (see Fig.~\ref{fig:sys}) in the left panel, while we only plot the distortion attained by the informed decoder, which has access to side information, in the right panel. The 1.53 dB refers to the mean-squared error gap that the entropy-constrained one-shot lattice quantizer is subjected to in high-rate regime~\cite{quantization}.}
\vspace{-0.5cm}
\label{fig:RD}
\end{figure*}

Following the popular class of neural compressors~\cite{Balle2017, balle2018variational, balle_journal}, we use stochastic gradient descent (SGD) to jointly optimize all learnable parameters. Since SGD relies on a Monte Carlo approximation of the expectations in the loss functions, we use the Gumbel-max technique, as in previous work~\cite{ozyilkan2023learned}, for sampling from discrete distributions. Similarly, we also leverage Concrete distributions~\cite{concrete} to aid stochastic optimization. Additional details about our experimental setup are provided in Appendix~\ref{sec:appendix_further_exp_setup}.

\section{Discussion}
\label{sec:discussion}

We first consider a setting where the proposed neural compressors clearly recover some elements of the optimal theoretical solution for the lossy source coding problem where side information may be absent. The visualization of the learned compressor obtained with the marginal scheme (see Fig.~\ref{fig:marginal_model}) is provided in Fig.~\ref{fig:visualizations}. As seen in the left panel of the figure, the neural encoder $p_\myVec{\omega}$ quantizes the source in a manner similar to standard point-to-point lossy compression. Looking at the right panel of the figure, we remark that the discontiguous quantization bins are learned as the neural compressor exhibits periodic-like mappings in the source space, akin to the binning-like behavior recovered by the neural WZ compressor proposed in~\cite{ozyilkan2023learned, ozyilkan2023neural, ozyilkan2023neuralworkshop}. We note that this grouping behavior is aligned with the achievability part of the HB theorem, which yields a rate discount similar to the random binning argument invoked in the WZ proof. The figure demonstrates that this learned distributed compressor, formulated based on the proposed marginal approach, exhibits a greater adaptability to robust scenarios compared to the WZ compression case previously explored in~\cite{ozyilkan2023learned, ozyilkan2023neural, ozyilkan2023neuralworkshop}. This is evident from the model's ability to recover both WZ and point-to-point coding strategies, despite not imposing any explicit structure exploiting the source knowledge onto the design. Although we noticed similar behaviors persisting in different experimental configurations other than the one considered in Fig.~\ref{fig:visualizations} for the marginal variant, we did not observe such consistent binning-like behavior in the visualizations of the compressors obtained by the joint and conditional formulations (not shown). For the conditional scheme, similar to the explanation in~\cite{ozyilkan2023learned}, we posit that this is due to having an ideal SW coder in the system design (see Fig.~\ref{fig:conditional_model}), unlike the marginal variant (see Fig.~\ref{fig:marginal_model}). We speculate that this choice of entropy coding scheme incentivizes the model to delegate the task of binning solely to the ideal SW code, rather than to the learned encoder.

In Fig.~\ref{subfig:RD_var0.1}, we provide R-D performances achieved with these three different formulations. The joint and marginal variants achieve similar performances, whereas the conditional model outperforms them, approaching the asymptotic HB bound. We attribute the comparable performances of the marginal and joint schemes to the operational equivalence between having two layered descriptions managed by separate classical entropy coders (see Eq.~\eqref{eq:upper_bound_sep_marg}) and having a single classical entropy coder with the input being the unified description of these two representations (see Eq.~\eqref{eq:upper_bound_merg}). We explain the improved R-D performance of the conditional formulation as follows. The SW code that this variant employs, which may leverage high-dimensional channel codes (e.g., as in DISCUS~\cite{DISCUS}), enables binning over long sequences, i.e., in a multi-shot fashion. This type of \emph{compress-bin}~\cite{network_info_theo} is much more efficient than the one that could be attained by the learned encoder, which could only bin a one-shot manner as it compresses each source realization one at a time. We refer the reader to Appendix~\ref{sec:appendix_experiments} for a discussion on an additional set of experiments.

In Fig.\ref{subfig:RD_ablation_var_0.1}, we illustrate the trade-off between system robustness and compression efficiency by evaluating distortion attained only with the informed decoders of all three proposed schemes, along with the learned WZ compressor proposed in~\cite{ozyilkan2023learned} (which operationally coincides with having $\beta=0$ in Eq.~\eqref{eq:expdist}, that is decoder-only side information is always assumed to be available). Recalling that all three proposed learned compressors are now also formulated to accommodate scenarios where side information may be absent, which is not the case for the learned WZ compressor studied in~\cite{ozyilkan2023learned}, a decline in distortion is expected. Interestingly, the conditional variant again surpasses all other learned schemes considered, underscoring the efficiency of high-dimensional binning capability facilitated by the SW coder.

\section{Conclusion} \label{sec:conclusion}
In this work, we have proposed three learning-based solutions to the problem of lossy source coding when side information may be absent, for which the optimal theoretical solution is asymptotic and non-constructive. By explicitly visualizing the behavior of the learned encoders and decoders, we demonstrated that they recover schemes aligned with the characteristics of the achievability of the HB theorem, that is WZ and standard lossy source coding. Future research directions include analyzing the robustness in fully distributed compression scenarios. Another interesting direction is to explore the performance of the models for more complex, high-dimensional sources such as images.

% \newpage

% \IEEEtriggeratref{12}
\bibliographystyle{IEEEtran}

\bibliography{ref.bib}

\providecommand{\noopsort}[1]{}
\begin{thebibliography}{10}
\providecommand{\url}[1]{#1}
\csname url@samestyle\endcsname
\providecommand{\newblock}{\relax}
\providecommand{\bibinfo}[2]{#2}
\providecommand{\BIBentrySTDinterwordspacing}{\spaceskip=0pt\relax}
\providecommand{\BIBentryALTinterwordstretchfactor}{4}
\providecommand{\BIBentryALTinterwordspacing}{\spaceskip=\fontdimen2\font plus
\BIBentryALTinterwordstretchfactor\fontdimen3\font minus \fontdimen4\font\relax}
\providecommand{\BIBforeignlanguage}[2]{{%
\expandafter\ifx\csname l@#1\endcsname\relax
\typeout{** WARNING: IEEEtran.bst: No hyphenation pattern has been}%
\typeout{** loaded for the language `#1'. Using the pattern for}%
\typeout{** the default language instead.}%
\else
\language=\csname l@#1\endcsname
\fi
#2}}
\providecommand{\BIBdecl}{\relax}
\BIBdecl

\bibitem{Wyner:IT:76}
A.~Wyner and J.~Ziv, ``The rate--distortion function for source coding with side information at the decoder,'' \emph{IEEE Transactions on Information Theory}, vol.~22, no.~1, pp. 1 -- 10, 1976.

\bibitem{elements_of_information_theory}
T.~M. Cover and J.~A. Thomas, \emph{Elements of Information Theory (Wiley Series in Telecommunications and Signal Processing)}.\hskip 1em plus 0.5em minus 0.4em\relax USA: Wiley-Interscience, 2006.

\bibitem{ozyilkan2023learned}
E.~Özyılkan, J.~Ballé, and E.~Erkip, ``Learned {W}yner–{Z}iv compressors recover binning,'' in \emph{2023 IEEE International Symposium on Information Theory (ISIT)}, 2023, pp. 701--706.

\bibitem{ozyilkan2023neural}
E.~Ozyilkan, J.~Ballé, and E.~Erkip, ``Neural distributed compressor discovers binning,'' \emph{IEEE Journal on Selected Areas in Information Theory}, pp. 1--1, 2024.

\bibitem{Mital_2022}
N.~Mital, E.~Özyılkan, A.~Garjani, and D.~Gündüz, ``Neural distributed image compression using common information,'' in \emph{2022 Data Compression Conference (DCC)}, 2022, pp. 182--191.

\bibitem{Mital_2023}
------, ``Neural distributed image compression with cross-attention feature alignment,'' in \emph{2023 IEEE/CVF Winter Conference on Applications of Computer Vision (WACV)}, 2023, pp. 2497--2506.

\bibitem{Li_2023}
\BIBentryALTinterwordspacing
P.~han Li, S.~K. Ankireddy, R.~Zhao, H.~N. Mahjoub, E.~M. Pari, ufuk topcu, S.~P. Chinchali, and H.~Kim, ``Task-aware distributed source coding under dynamic bandwidth,'' in \emph{Thirty-seventh Conference on Neural Information Processing Systems}, 2023. [Online]. Available: \url{https://openreview.net/forum?id=1A4ZqTmnye}
\BIBentrySTDinterwordspacing

\bibitem{Heegard_Berger85}
C.~Heegard and T.~Berger, ``Rate distortion when side information may be absent,'' \emph{IEEE Transactions on Information Theory}, vol.~31, no.~6, pp. 727--734, 1985.

\bibitem{Kaspi}
A.~Kaspi, ``Rate-distortion function when side-information may be present at the decoder,'' \emph{IEEE Transactions on Information Theory}, vol.~40, no.~6, pp. 2031--2034, 1994.

\bibitem{Ishwar}
P.~Ishwar, R.~Puri, S.~Pradhan, and K.~Ramchandran, ``On compression for robust estimation in sensor networks,'' in \emph{IEEE International Symposium on Information Theory, 2003. Proceedings.}, 2003, pp. 193--.

\bibitem{ozyilkan2023neuralworkshop}
E.~Ozyilkan, J.~Ball{\'e}, and E.~Erkip, ``Neural distributed compressor does binning,'' in \emph{ICML 2023 Workshop Neural Compression: From Information Theory to Applications}, 2023.

\bibitem{DISCUS}
S.~Pradhan and K.~Ramchandran, ``Distributed source coding using syndromes ({DISCUS}): {D}esign and construction,'' \emph{IEEE Transactions on Information Theory}, vol.~49, no.~3, pp. 626--643, 2003.

\bibitem{hornik_et_al}
K.~Hornik, M.~Stinchcombe, and H.~White, ``Multilayer feedforward networks are universal approximators,'' \emph{Neural Networks}, vol.~2, no.~5, p. 359–366, jul 1989.

\bibitem{Leshno1993}
\BIBentryALTinterwordspacing
M.~Leshno, V.~Y. Lin, A.~Pinkus, and S.~Schocken, ``Multilayer feedforward networks with a nonpolynomial activation function can approximate any function,'' \emph{Neural Networks}, vol.~6, no.~6, pp. 861--867, Jan. 1993. [Online]. Available: \url{https://doi.org/10.1016/s0893-6080(05)80131-5}
\BIBentrySTDinterwordspacing

\bibitem{Balle2017}
J.~Ball\'{e}, V.~Laparra, and E.~P. Simoncelli, ``End-to-end optimized image compression,'' in \emph{International Conference on Learning Representations (ICLR)}, 2017.

\bibitem{balle2018variational}
J.~Ballé, D.~Minnen, S.~Singh, S.~J. Hwang, and N.~Johnston, ``Variational image compression with a scale hyperprior,'' 2018.

\bibitem{Chris_Andrea_2012}
C.~T.~K. Ng, C.~Tian, A.~J. Goldsmith, and S.~Shamai, ``Minimum expected distortion in gaussian source coding with fading side information,'' \emph{IEEE Transactions on Information Theory}, vol.~58, no.~9, pp. 5725--5739, 2012.

\bibitem{kullback1997information}
S.~Kullback, \emph{Information Theory and Statistics}, ser. A Wiley Publication in Mathematical Statistics.\hskip 1em plus 0.5em minus 0.4em\relax Dover Publications, 1997.

\bibitem{quantization}
R.~Gray and D.~Neuhoff, ``Quantization,'' \emph{IEEE Transactions on Information Theory}, vol.~44, no.~6, pp. 2325--2383, 1998.

\bibitem{balle_journal}
J.~Ballé, P.~A. Chou, D.~Minnen, S.~Singh, N.~Johnston, E.~Agustsson, S.~J. Hwang, and G.~Toderici, ``Nonlinear transform coding,'' \emph{IEEE Journal of Selected Topics in Signal Processing}, vol.~15, no.~2, pp. 339--353, 2021.

\bibitem{concrete}
\BIBentryALTinterwordspacing
C.~J. Maddison, A.~Mnih, and Y.~W. Teh, ``The concrete distribution: A continuous relaxation of discrete random variables,'' in \emph{International Conference on Learning Representations}, 2017. [Online]. Available: \url{https://openreview.net/forum?id=S1jE5L5gl}
\BIBentrySTDinterwordspacing

\bibitem{network_info_theo}
A.~E. Gamal and Y.-H. Kim, \emph{Network Information Theory}.\hskip 1em plus 0.5em minus 0.4em\relax USA: Cambridge University Press, 2012.

\bibitem{jax}
\BIBentryALTinterwordspacing
J.~Bradbury, R.~Frostig, P.~Hawkins, M.~J. Johnson, C.~Leary, D.~Maclaurin, G.~Necula, A.~Paszke, J.~Vander{P}las, S.~Wanderman-{M}ilne, and Q.~Zhang, ``{JAX}: Composable transformations of {P}ython+{N}um{P}y programs,'' 2018. [Online]. Available: \url{http://github.com/google/jax}
\BIBentrySTDinterwordspacing

\bibitem{adam}
\BIBentryALTinterwordspacing
D.~P. Kingma and J.~Ba, ``Adam: A method for stochastic optimization,'' 2014. [Online]. Available: \url{https://arxiv.org/abs/1412.6980}
\BIBentrySTDinterwordspacing

\end{thebibliography}

\newpage
\onecolumn
\appendices

\section{The Minimum Distortion For Eq.~\eqref{eq:expdist}} \label{sec:appendix_minimum_distortion}

When considering the weighted distortion metric as in Eq. \eqref{eq:expdist} for the quadratic-Gaussian setup of the HB problem, the minimum distortion is provided by \cite{Chris_Andrea_2012}:
\begin{align}
\label{eq:min_weighted_dist}
    \myVec{D}^*(R) = \beta D_1^* + (1-\beta)D_2^*,
\end{align}
where
\begin{align*}
    &D_1^* = \Bigg[\left(2^{-2R}\sigma_x^2\sigma_n^2\frac{1-\beta}{\beta}\right)^{1/2} - \sigma_n^2 \Bigg]_{[D_1^-,D_1^+]}, \\
    &D_2^* =  \Bigg[ \left(2^{-2R}\sigma_x^2\sigma_n^2\frac{\beta}{1-\beta}\right)^{1/2} \Bigg]_{[D_2^-,D_2^+]}.
\end{align*}
Here, we used the notation $[x]_{[a,b]} = \min(\max(a,x),b)$, and
\begin{align*}
    D_1^- &=  2^{-2R}\sigma_x^2  \qquad \quad \qquad \;\;, \; D_1^+ = \sigma_x^2, \\
    D_2^- &=  2^{-2R}\left(\frac{1}{\sigma_x^2}+\frac{1}{\sigma_n^2}\right)^{-1} \;, \; D_2^+ = \left(\frac{2^{2R}}{\sigma_x^2}+\frac{1}{\sigma_n^2}\right)^{-1}.
\end{align*}
To further explain the above expressions, $D_1^-$ represents the minimum achievable distortion for $\hat{X}_1$ when allocating the entire rate $R$ for the description of $W$, while $D_1^+$ denotes the worst-case distortion attained for $\hat{X}_1$ when the entire rate is dedicated to the description of $U$. Likewise, $D_2^-$ and $D_2^+$ are the minimum and maximum distortions for $\hat{X}_2$ achieved when the rate $R$ is utilized for $U$ and $W$, respectively.

\section{Proof of the Upper Bound in Eq.~\eqref{eq:joint_upper_bound}} \label{sec:appendix_upper_bound}

Building onto Theorem~\ref{theo:HB}, we can re-write the second mutual information term in Eq.~\eqref{eq:HB} as:
\begin{align}
    I(X;U \vert Y, W) &= I(U;X,Y \vert W) - I(U;Y \vert W), \label{eq:(a)} \\
    &= I(U;X \vert W) - I(U; Y \vert W) \label{eq:(b)},
\end{align}
where Eq.~\eqref{eq:(a)} follows from the chain rule for mutual information, and Eq.~\eqref{eq:(b)} is due to the Markov chain $(W,U) - X - Y$. Then, Eq.~\eqref{eq:HB} becomes:
\begin{align}
    I(X;W) + I(X;U \vert Y, W) & = I(X;W) + I(U;X \vert W) - I(U; Y \vert W),  \nonumber  \\
    & \leq I(X;W) + I(U;X\vert W), \label{eq:(c)} \\
    & = I(X;W,U), \label{eq:(d)} \\
    & = I(X;V), \label{eq:(e)}
\end{align}
where Eq.~\eqref{eq:(c)} is due to the non-negativity of mutual information, Eq.~\eqref{eq:(d)} follows from the chain rule for mutual information, and Eq.~\eqref{eq:(e)} is due to the definition of $V = (W, U)$.

\section{Operational Heegard--Berger Compressors Illustrated in Figs.~\ref{fig:unified_model},~\ref{fig:marginal_model} and~\ref{fig:conditional_model}} \label{sec:appendix_operational}

The upper bound in Eq.~\eqref{eq:upper_bound_merg} corresponds to the compression rate of a system employing a one-shot encoder in the \emph{joint} model and an entropy code which asymptotically achieves a cross-entropy rate of
\begin{align}
\mathbb{E}_{p(x)} [\mathbb{E}_{v \sim p_{\boldsymbol{\theta}}(v \vert x)}[-\log q_{\boldsymbol{\eta}}(v)]].
\end{align}
Similarly, the upper bounds in Eqs.~\eqref{eq:upper_bound_sep_marg} and~\eqref{eq:upper_bound_sep_cond} coincide with the compression rates of designs having one-shot compressors coupled with either classic entropy coders (i.e., the \emph{marginal} model) or a combination of a classic entropy coder and an ideal SW coder (i.e., the \emph{conditional} model), respectively, which asymptotically achieve the cross-entropy rates of
\begin{align}
\mathbb{E}_{p(x)}[\mathbb{E}_{w\sim p_{\boldsymbol{\omega}}(w\vert x), u \sim p_{\boldsymbol{\gamma}}(u\vert w, x)}[-\log q_{\boldsymbol{\zeta}}(w) - \log q_{\boldsymbol{\psi}}(u \vert w)]],
\end{align} and
\begin{align}
\mathbb{E}_{p(x, y)}[\mathbb{E}_{w\sim p_{\boldsymbol{\omega}}(w\vert x), u \sim p_{\boldsymbol{\gamma}}(u\vert w, x)}[-\log q_{\boldsymbol{\zeta}}(w) - \log q_{\boldsymbol{\mu}}(u \vert w, y)]].
\end{align} Therefore, by minimizing $L_{\mathrm{j}}, L_{\mathrm{m}},$ and $L_{c}$ in Eqs.~\eqref{eq:proposed_loss_joint},~\eqref{eq:proposed_loss_marginal} and~\eqref{eq:proposed_loss_conditional}, we optimize the robust operational distributed compression schemes depicted in Figs.~\ref{fig:unified_model},~\ref{fig:marginal_model} and~\ref{fig:conditional_model}, respectively, in an end-to-end fashion. We remark that unlike the neural WZ compressors proposed in~\cite{ozyilkan2023learned, ozyilkan2023neural, ozyilkan2023neuralworkshop}, the proposed distributed compressors illustrated in Fig.~\ref{fig:sys} can by design accommodate for scenarios where decoder-only side information may be absent. 

\section{Additional Details on Experimental Setup} \label{sec:appendix_further_exp_setup}

For the probabilistic models conditioned on other variables (e.g., 
$ p_{\boldsymbol{\theta}}, q_{\boldsymbol{\psi}}$) and for the decoding functions (e.g., $g_{\boldsymbol{\phi}}, g_{\boldsymbol{\kappa}}$), we employ ANNs with three dense layers. Each layer, except the last one, consists of 100 units followed by a leaky rectified linear unit as the activation function. As for the other probabilistic model (i.e., $q_{\boldsymbol{\eta}}$ and $q_{\boldsymbol{\zeta}}$), we directly treat them as learnable parameters. In our experiments, we observed that increasing the dimensions of ANNs or employing different activation functions did not lead to better performance. For probabilistic models and functions that have more than one input, we feed ANNs with concatenated version of the inputs. We conduct our experiments using the JAX~\cite{jax} framework. We train all learning-based compressors for 500 epochs with randomly initialized network weights and utilize Adam~\cite{adam}, a widely used variant of SGD. 
We employ a learning rate of \( 1 \times 10^{-4} \) and a batch size of \( B = 1024 \), which aligns with the number of realizations sampled from the aforementioned correlation model in Section~\ref{subsec:experimental_setup}. Similar to~\cite{ozyilkan2023learned}, we likewise choose $q_{\boldsymbol{\eta}}(v), q_{\boldsymbol{\zeta}}(w), q_{\boldsymbol{\psi}}{(u \vert w)}$ and $q_{\boldsymbol{\mu}}{(u \vert w, y)}$ as Concrete during training to align the distributions of $v, w$ and $u$ samples.
 
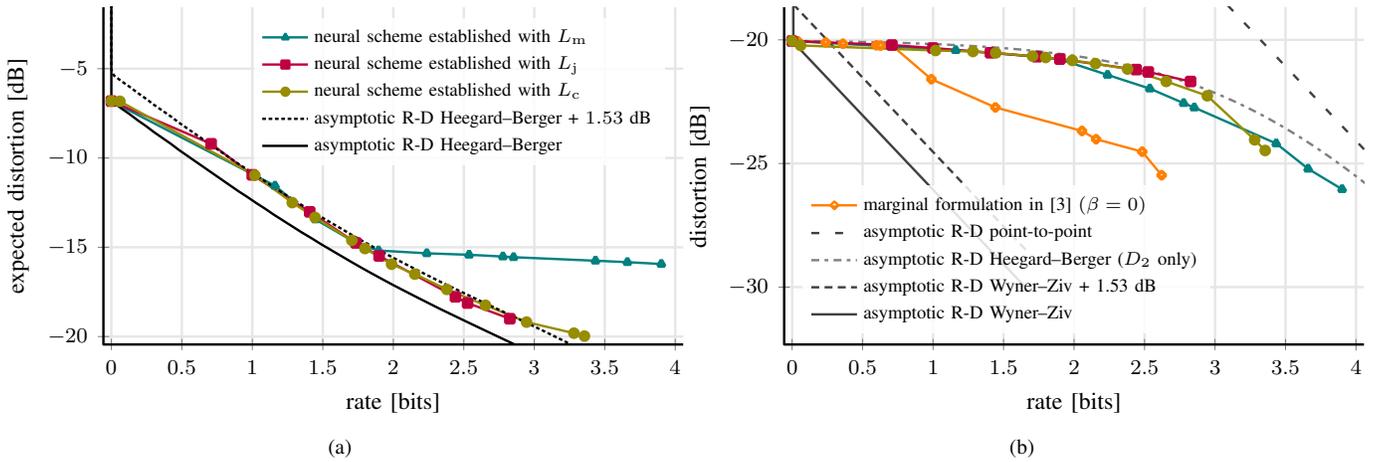
\begin{figure}
\begin{subfigure}{0.5\columnwidth}
    \raggedleft
    \begin{tikzpicture}[trim axis right]
    \begin{axis}[
      height=.19\textheight,
      width=.85\linewidth,
      scale only axis,
      xlabel={rate [bits]},
      ylabel={expected distortion [dB]},
      xmin=0.,
      xmax=4.,
      ymin=-20.,
      ymax=-2.,
      legend pos=north east,
      legend style={font=\scriptsize},
      ]
      \addplot[color=teal, mark=triangle*] table {
    0.019437 -6.902399894202913
    1.160632 -11.555103982666521
    1.448439 -13.442278809118509
    1.750403 -14.825137894202772
    1.895083 -15.1818486936258
    2.237641 -15.3384443524779
    2.537413 -15.421029713177159
    2.77756 -15.518659037352215
    2.8528888 -15.556398458452044
    3.43418 -15.75461362845946
    3.660454 -15.833761326158161
    3.901154 -15.93766488625677
      };
       \addplot[color=purple,mark=square*] table {
        0.000108 -6.81556823522012
        0.707603 -9.206342180086773
        0.999261 -10.942634682746892
        1.406978 -13.018385355757582
        1.73565 -14.756947711550765
        1.899453 -15.486741915104805
        2.440638 -17.77491623960329
        2.526866 -18.122537302195408
        2.826519 -19.001812903847696
      };
      \addplot[color=olive,mark=*] table {
        2e-05 -6.809085215650757
        0.060256 -6.824986882298125
        1.017603 -10.967200507234443
        1.282936 -12.500087464978701
        1.44445 -13.327712569662925
        1.705718 -14.613383103615963
        1.798754 -15.071418708028874
        1.987731 -15.947902856153666
        2.152254 -16.504310974913242
        2.379484 -17.35843563254114
        2.653782 -18.249232788237897
        2.945144 -19.190850411433747
        3.281108 -19.811331368490933
        3.355886 -19.974020192800914
      };
      \addplot[color=black,densely dotted] table {
4.924341103082452 -28.47
4.8649195957105 -28.14
4.805036551765327 -27.810000000000002
4.744660038683169 -27.479999999999997
4.6837567282366575 -27.15
4.622291955660514 -26.82
4.560841177875604 -26.49
4.498274265000788 -26.16
4.434856863560662 -25.83
4.371376050761171 -25.499999999999996
4.307040605587071 -25.169999999999998
4.241799845516215 -24.84
4.175602948801351 -24.509999999999998
4.108399394748775 -24.179999999999996
4.0401394475274195 -23.849999999999998
3.9711163492923514 -23.519999999999996
3.9008242569206306 -23.189999999999998
3.8292120184151557 -22.86
3.7562330467369636 -22.529999999999994
3.681979763355857 -22.200000000000003
3.606184870575351 -21.869999999999997
3.5288252299358116 -21.54
3.4498892784566237 -21.21
3.369383322702206 -20.88
3.2872936347973214 -20.55
3.203679464952611 -20.22
3.1186950769767905 -19.889999999999997
3.032346941409026 -19.56
2.945036055252714 -19.23
2.8569272864248374 -18.9
2.7684948742791278 -18.569999999999997
2.6803228719526473 -18.24
2.5919873270007936 -17.909999999999997
2.5043464162599496 -17.58
2.4176992674590223 -17.249999999999996
2.3308665441240333 -16.919999999999998
2.247782645405534 -16.589999999999996
2.1650888880402177 -16.259999999999998
2.0833472047086095 -15.930000000000001
2.0023110258475443 -15.6
1.9243929453682491 -15.270000000000001
1.847656406376351 -14.94
1.7725107628574097 -14.610000000000001
1.6996645097491192 -14.28
1.628409152113785 -13.950000000000001
1.5571537944784513 -13.62
1.4858984368431172 -13.290000000000001
1.4191383543076925 -12.959999999999999
1.3488689029290135 -12.63
1.2830947266502437 -12.3
1.2173205503714737 -11.969999999999999
1.1515463740927039 -11.64
1.0905102792072616 -11.309999999999999
1.0254792028917286 -10.979999999999999
0.9641021969897765 -10.65
0.8994120316907535 -10.319999999999999
0.8391190367685475 -9.99
0.7788260418463422 -9.66
0.7185330469241363 -9.33
0.658240052001931 -9.0
0.597947057079725 -8.67
0.5376540621575197 -8.340000000000002
0.47736106723531385 -8.009999999999998
0.4225492536696725 -7.679999999999997
0.36225625874746714 -7.349999999999999
0.30196326382526123 -7.0200000000000005
0.2471514502596198 -6.69
0.18685845533741383 -6.359999999999997
0.13204664177177242 -6.029999999999998
0.07175364684956725 -5.699999999999998
0.016941833283925876 -5.369999999999999
0.0 -5.04
0.0 -4.709999999999996
0.0 -4.379999999999997
0.0 -4.049999999999998
0.0 -3.719999999999999
0.0 -3.3899999999999997
0.0 -3.0600000000000005
0.0 -2.729999999999997
0.0 -2.3999999999999977
0.0 -2.0699999999999985
0.0 -1.7399999999999995
0.0 -1.4100000000000004
0.0 -1.0799999999999967
0.0 -0.749999999999998
0.0 -0.41999999999999815
0.0 -0.08999999999999941
0.0 0.24
0.0 0.5699999999999993
0.0 0.9000000000000026
0.0 1.230000000000002
0.0 1.560000000000001
0.0 1.8900000000000001
0.0 2.2199999999999993
0.0 2.5500000000000025
0.0 2.880000000000002
0.0 3.2100000000000013
0.0 3.540000000000001
0.0 3.87
0.0 4.200000000000004
0.0 4.529999999999999
      };
      \addplot[color=black] table {
      4.924341103082452 -30.0
4.8649195957105 -29.67
4.805036551765327 -29.340000000000003
4.744660038683169 -29.009999999999998
4.6837567282366575 -28.68
4.622291955660514 -28.35
4.560841177875604 -28.02
4.498274265000788 -27.69
4.434856863560662 -27.36
4.371376050761171 -27.029999999999998
4.307040605587071 -26.7
4.241799845516215 -26.37
4.175602948801351 -26.04
4.108399394748775 -25.709999999999997
4.0401394475274195 -25.38
3.9711163492923514 -25.049999999999997
3.9008242569206306 -24.72
3.8292120184151557 -24.39
3.7562330467369636 -24.059999999999995
3.681979763355857 -23.730000000000004
3.606184870575351 -23.4
3.5288252299358116 -23.07
3.4498892784566237 -22.740000000000002
3.369383322702206 -22.41
3.2872936347973214 -22.080000000000002
3.203679464952611 -21.75
3.1186950769767905 -21.419999999999998
3.032346941409026 -21.09
2.945036055252714 -20.76
2.8569272864248374 -20.43
2.7684948742791278 -20.099999999999998
2.6803228719526473 -19.77
2.5919873270007936 -19.439999999999998
2.5043464162599496 -19.11
2.4176992674590223 -18.779999999999998
2.3308665441240333 -18.45
2.247782645405534 -18.119999999999997
2.1650888880402177 -17.79
2.0833472047086095 -17.46
2.0023110258475443 -17.13
1.9243929453682491 -16.8
1.847656406376351 -16.47
1.7725107628574097 -16.14
1.6996645097491192 -15.809999999999999
1.628409152113785 -15.48
1.5571537944784513 -15.149999999999999
1.4858984368431172 -14.82
1.4191383543076925 -14.489999999999998
1.3488689029290135 -14.16
1.2830947266502437 -13.83
1.2173205503714737 -13.499999999999998
1.1515463740927039 -13.17
1.0905102792072616 -12.839999999999998
1.0254792028917286 -12.509999999999998
0.9641021969897765 -12.18
0.8994120316907535 -11.849999999999998
0.8391190367685475 -11.52
0.7788260418463422 -11.19
0.7185330469241363 -10.86
0.658240052001931 -10.53
0.597947057079725 -10.2
0.5376540621575197 -9.870000000000001
0.47736106723531385 -9.539999999999997
0.4225492536696725 -9.209999999999997
0.36225625874746714 -8.879999999999999
0.30196326382526123 -8.55
0.2471514502596198 -8.22
0.18685845533741383 -7.889999999999997
0.13204664177177242 -7.559999999999998
0.07175364684956725 -7.229999999999999
0.016941833283925876 -6.8999999999999995
0.0 -6.57
0.0 -6.239999999999997
0.0 -5.9099999999999975
0.0 -5.579999999999998
0.0 -5.249999999999999
0.0 -4.92
0.0 -4.590000000000001
0.0 -4.259999999999997
0.0 -3.929999999999998
0.0 -3.5999999999999988
0.0 -3.2699999999999996
0.0 -2.9400000000000004
0.0 -2.6099999999999968
0.0 -2.279999999999998
0.0 -1.9499999999999982
0.0 -1.6199999999999994
0.0 -1.29
0.0 -0.9600000000000007
0.0 -0.6299999999999975
0.0 -0.2999999999999981
0.0 0.030000000000001026
0.0 0.36000000000000004
0.0 0.6899999999999994
0.0 1.0200000000000027
0.0 1.350000000000002
0.0 1.6800000000000013
0.0 2.0100000000000007
0.0 2.3400000000000003
0.0 2.6700000000000035
0.0 2.999999999999999
      };
      \legend{neural scheme established with $L_{\mathrm{m}}$, neural scheme established with $L_{\mathrm{j}}$, neural scheme established with $L_{\mathrm{c}}$,
      asymptotic R-D Heegard--Berger + $1.53$ dB, asymptotic R-D Heegard--Berger}
    \end{axis}
    \end{tikzpicture}
  \caption{}
  \label{subfig:RD_var0.01}
\end{subfigure}%
\hfill%
\begin{subfigure}{0.5\columnwidth}
    \raggedleft
    \begin{tikzpicture}[trim axis right]
    \begin{axis}[
      height=.19\textheight,
      width=.85\linewidth,
      scale only axis,
      xlabel={rate [bits]},
      ylabel={distortion [dB]},
      xmin=0.,
      xmax=4.,
      ymin=-32.,
      ymax=-19.,
      legend pos=south west,
      legend style={font=\scriptsize},
      ]
    \addplot[color=orange,mark=halfsquare*] table {
    6.1e-05 -20.04233220950026
0.0024 -20.043648054024498
0.00731 -20.040578370074495
0.041781 -20.057708591823015
0.238379 -20.10594681998484
0.360778 -20.15742798294584
0.598252 -20.2200219004126
0.627762 -20.232374767325393
0.709311 -20.229626647753186
0.98805 -21.592667653881932
1.44228 -22.732727909734276
2.056118 -23.68454772165691
2.156501 -24.01866354186762
2.483355 -24.523483416400307
2.621599 -25.476001540885584
      };
    \addplot[color=darkgray,loosely dashed] table {
      4.9828921423310435 -30.0
      0.049828921423310135 -0.2999999999999981
      0.0 0.030000000000001026
      };
    \addplot[color=gray,dashdotted] table {
0	-20.0432137378264
0.0505050505050505	-20.0463311165239
0.101010101010101	-20.0496720952816
0.151515151515152	-20.0532525213690
0.202020202020202	-20.0570893368978
0.252525252525253	-20.0612006502708
0.303030303030303	-20.0656058116701
0.353535353535354	-20.0703254927184
0.404040404040404	-20.0753817704348
0.454545454545455	-20.0807982155910
0.505050505050505	-20.0865999855564
0.555555555555556	-20.0928139216953
0.606060606060606	-20.0994686513507
0.656565656565657	-20.1065946944157
0.707070707070707	-20.1142245744509
0.757575757575758	-20.1223929342583
0.808080808080808	-20.1311366557645
0.858585858585859	-20.1404949840008
0.909090909090909	-20.1505096548920
0.959595959595960	-20.1612250264771
1.01010101010101	-20.1726882130890
1.06060606060606	-20.1849492219074
1.11111111111111	-20.1980610911751
1.16161616161616	-20.2120800292295
1.21212121212121	-20.2270655533481
1.26262626262626	-20.2430806272380
1.31313131313131	-20.2601917958198
1.36363636363636	-20.2784693157564
1.41414141414141	-20.2979872799696
1.46464646464646	-20.3188237341655
1.51515151515152	-20.3410607831580
1.56565656565657	-20.3647846845426
1.61616161616162	-20.3900859270328
1.66666666666667	-20.4170592905326
1.71717171717172	-20.4458038847910
1.76767676767677	-20.4764231632695
1.81818181818182	-20.5090249086661
1.86868686868687	-20.5437211863832
1.91919191919192	-20.5806282621221
1.96969696969697	-20.6198664797300
2.02020202020202	-20.6615600954526
2.07070707070707	-20.7058370648460
2.12121212121212	-20.7528287788083
2.17171717171717	-20.8026697455088
2.22222222222222	-20.8554972154336
2.27272727272727	-20.9114507473456
2.32323232323232	-20.9706717136789
2.37373737373737	-21.0333027447545
2.42424242424242	-21.0994871122234
2.47474747474748	-21.1693680532950
2.52525252525253	-21.2430880385937
2.57575757575758	-21.3207879878720
2.62626262626263	-21.4026064392712
2.67676767676768	-21.4886786793244
2.72727272727273	-21.5791358423928
2.77777777777778	-21.6741039896671
2.82828282828283	-21.7737031791935
2.87878787878788	-21.8780465395435
2.92929292929293	-21.9872393606700
2.97979797979798	-22.1013782161385
3.03030303030303	-22.2205501312225
3.08080808080808	-22.3448318112897
3.13131313131313	-22.4742889444289
3.18181818181818	-22.6089755913883
3.23232323232323	-22.7489336746055
3.28282828282828	-22.8941925764371
3.33333333333333	-23.0447688546900
3.38383838383838	-23.2006660812702
3.43434343434343	-23.3618748072771
3.48484848484849	-23.5283726552649
3.53535353535354	-23.7001245367650
3.58585858585859	-23.8770829905955
3.63636363636364	-24.0591886350829
3.68686868686869	-24.2463707251467
3.73737373737374	-24.4385478033399
3.78787878787879	-24.6356284324341
3.83838383838384	-24.8375119960289
3.88888888888889	-25.0440895529697
3.93939393939394	-25.2552447310723
3.98989898989899	-25.4708546457541
4.04040404040404	-25.6907908296377
4.09090909090909	-25.9149201599689
4.14141414141414	-26.1431057717288
4.19191919191919	-26.3752079455626
4.24242424242424	-26.6110849610328
4.29292929292929	-26.8505939071791
4.34343434343434	-27.0935914438707
4.39393939393939	-27.3399345089274
4.44444444444445	-27.5894809674198
4.49494949494950	-27.8420902008992
4.54545454545455	-28.0976236355362
4.59595959595960	-28.3559452092413
4.64646464646465	-28.6169217787884
4.69696969696970	-28.8804234687678
4.74747474747475	-29.1463239648478
4.79797979797980	-29.4145007543419
4.84848484848485	-29.6848353174632
4.89898989898990	-29.9572132729098
4.94949494949495	-30.2315244815866
5	-30.5076631123304
      };
    \addplot[color=darkgray,densely dashed] table {
      1.6537864009551462 -28.47
      0.009431993985901525 -18.569999999999997
      0.0 -18.24
      };
    \addplot[color=darkgray] table {
        1.6537864009551462 -30.0
        0.009431993985901525 -20.099999999999998
        0.0 2.999999999999999
      };
    \addplot[color=teal, mark=triangle*] table {
    0.019437 -20.04979393875242
1.008507 -20.36637233896264
1.160632 -20.417706858116176
1.448439 -20.541868734126613
1.750403 -20.685929864434268
1.895083 -20.779495978328264
2.237641 -21.41763664570487
2.537413 -21.972262747080244
2.77756 -22.560201347581568
2.8528888 -22.739047619979328
3.43418 -24.19189027339054
3.660454 -25.215778122599197
3.901154 -26.041496239812187
         };
     \addplot[color=purple,mark=square*] table {
0.000108 -20.047158923107403
0.707603 -20.20998251525279
0.999261 -20.32967117841298
1.406978 -20.523702526156455
1.73565 -20.681862960408605
1.899453 -20.777937225609836
2.440638 -21.202735033604228
2.526866 -21.29888844635599
2.826519 -21.68450148004244
      };
      \addplot[color=olive,mark=*] table {
2e-05 -20.039701715889233
0.060256 -20.22688026603074
1.017603 -20.431115453499228
1.282936 -20.470139348029445
1.44445 -20.528114344739063
1.705718 -20.659057316445196
1.798754 -20.711946291067157
1.987731 -20.83177715404088
2.152254 -20.95988116402612
2.379484 -21.180450286603993
2.653782 -21.682581663543615
2.945144 -22.262866747229783
3.281108 -24.04393565134397
3.355886 -24.474534520443395
      };
      \legend{marginal formulation in~\cite{ozyilkan2023learned} $(\beta = 0)$, asymptotic R-D point-to-point, asymptotic R-D Heegard--Berger ($D_2$ only), asymptotic R-D Wyner--Ziv + 1.53 dB, asymptotic R-D Wyner--Ziv};
    \end{axis}
    \end{tikzpicture}
    \caption{}
    \label{subfig:RD_ablation_var_0.01}
\end{subfigure}%
\caption{Rate--distortion (R-D) performances obtained with joint, marginal and conditional formulations (see Eqs.~\eqref{eq:proposed_loss_joint},~\eqref{eq:proposed_loss_marginal} and~\eqref{eq:proposed_loss_conditional}), where experimental setup parameters (see Section~\ref{subsec:experimental_setup}) are set as $\sigma^{2}_{x}=1.00$, $\sigma^{2}_{n} = 0.01$ and $\beta = 0.20$. In both panels, we plot the empirical results versus the asymptotic bounds. We provide the expected distortion achieved by two decoders (see Fig.~\ref{fig:sys}) in the left panel, while we only plot the distortion attained by the informed decoder, which has access to side information, in the right panel. The 1.53 dB refers to the mean-squared error gap that the entropy-constrained one-shot lattice quantizer is subjected to in high-rate regime~\cite{quantization}.}
\label{fig:RD_var_0.1}
\end{figure}

\section{Additional Experiments} \label{sec:appendix_experiments}
 
Fig.~\ref{subfig:RD_var0.01} shows R-D results for all three different formulations we consider for the HB problem (i.e., \emph{joint}, \emph{marginal} and \emph{conditional} as explained in Section~\ref{sec:neural_upper_bounds}), where we set experimental setup parameters as $\sigma_x^2 = 1.00$, $\sigma_n^2 = 0.01$ and $\beta=0.20$ (see Fig.~\ref{fig:visualizations}). In the asymptotic blocklength regime, under this choice of correlation model parameters, the optimal R-D trade-off for the minimum expected distortion can be attained by operating in the point-to-point R-D region for all rate values (see Section~\ref{subsec:experimental_setup} and Appendix~\ref{sec:appendix_minimum_distortion} for the discussion on different operational R-D regions). This implies that the optimal coding strategy is to use the entire rate budget for the description of $W$. We remark that our learned distributed compressors emulate this optimal approach. The joint and the conditional models (nearly) allocate all of their rates to the description of $W$, and achieve R-D performance close to the asymptotic R-D bound. As in Fig.~\ref{subfig:RD_var0.1}, the compressors do not reach the asymptotic R-D bound as they operate in a one-shot quantization regime in tandem with variable rate entropy coding. We observe that the marginal model behaves similarly to the other schemes at low rates, but deviates from them at higher rate values. We speculate that this occurs because, at higher rates, the marginal model, employing a layered encoding strategy, tends to allocate some rate for the description of $U$, which may not be optimal given this specific choice of correlation model parameters.

Similar to the comparison shown in Fig.~\ref{subfig:RD_ablation_var_0.1}, we illustrate in Fig.\ref{subfig:RD_ablation_var_0.01} the trade-off between system robustness and compression efficiency by evaluating distortions attained only with the informed decoders of all three proposed schemes, along with the learned WZ compressor established in~\cite{ozyilkan2023learned} (which operationally coincides with having $\beta=0$ in Eq.~\eqref{eq:expdist}, that is this learned compressor always presume the presence of decoder-only side information). We observe that unlike the scenario depicted in Fig.\ref{subfig:RD_ablation_var_0.1}, all three variants exhibit inferior performance compared to this learned WZ compressor, primarily due to setting a higher $\beta$ value. Hence, given that all of the three learned schemes are trained to prioritize the uninformed decoder more than the experimental setup considered in Fig.~\ref{subfig:RD_ablation_var_0.1}, such a decrease in performance considering the WZ compression scenario is anticipated. Moreover, the decline in distortion at higher rates for the marginal formulation illustrated in Fig. \ref{subfig:RD_var0.01} is reflected as a minor R-D enhancement in Fig.~\ref{subfig:RD_ablation_var_0.01}. We hypothesize that this happens because, at higher rates, the marginal approach starts allocating more rates to the description of $U$. On the other hand, this marginal scheme also seeks to minimize this rate budget of $U$ by recovering periodic-like mappings, akin to binning, at higher rate values as observed in the right panel of Fig.~\ref{fig:visualizations}. This, in return, results in a slightly improved distortion performance attained by the informed decoder, as shown in Fig.~\ref{subfig:RD_ablation_var_0.01}. We remark that the HB curve in Fig.\ref{subfig:RD_ablation_var_0.01} shows the distortion attained by the informed decoder of the HB in Fig.\ref{subfig:RD_var0.01} which is optimal for expected distortion. Since our learned compressors are trained for minimizing the expected distortion as well, the HB curve in Fig.\ref{subfig:RD_ablation_var_0.01} is not necessarily a lower bound for the distortions attained by the informed decoders of our models.

\end{document}